\documentclass[3p, 10pt, onecolumn]{elsarticle}
\usepackage{graphicx,amscd,amsmath,amssymb,verbatim}
\usepackage{amsfonts,epsfig}
\usepackage{mathptmx}
\usepackage{setspace}
\usepackage{epsfig}
\usepackage{url}
\usepackage{multirow}
\usepackage{subfigure}
\usepackage{graphicx,color}
\usepackage[ruled,vlined]{algorithm2e}
\usepackage[normalem]{ulem}

\pdfoutput=1

\begin{document}


\journal{Elsevier}

\begin{frontmatter}

\title{A thorough study of the performance of simulated annealing in the traveling salesman problem under correlated and long tailed spatial scenarios}

\author{Roberto da Silva$^{1}$, Eliseu Venites Filho $^{1}$, Alexandre Alves $^{2}$}

\address{1-Instituto de F{\'i}sica, Universidade Federal do Rio Grande do Sul, UFRGS, Porto Alegre-RS, Brazil}

\address{2-Departamento de F\'{i}sica, Universidade Federal de S\~{a}o Paulo, UNIFESP, Diadema-SP, Brazil\\}


\begin{abstract}

Metaheuristics, as the simulated annealing used in the optimization of disordered systems, goes beyond physics, and the traveling salesman is a paradigmatic NP-complete problem that allows inferring important theoretical properties of the algorithm in different random environments. Many versions of the algorithm are explored in the literature, but so far the effects of the statistical distribution of the coordinates of the cities on the performance of the algorithm have been neglected. We propose a simple way to explore this aspect by analyzing the performance of a standard version of the simulated annealing (geometric cooling) in correlated systems with a simple and useful method based on a linear combination of independent random variables. Our results suggest that performance depends on the shape of the statistical distribution of the coordinates but not necessarily on its variance corroborated by the cases of uniform and normal distributions. On the other hand, a study with different power laws (different decay exponents) for the coordinates always produces different performances. We show that the performance of the simulated annealing, even in its best version, is not improved when the distribution of the coordinates does not have the first moment. However, surprisingly, we still observe improvements in situations where the second moment is not defined but not where the first one is not defined. Finite-size scaling fits, and universal laws support all of our results. In addition, our study show when the cost must be scaled.

\end{abstract}
\end{frontmatter}

\section{Introduction}

\label{sec:introduction}

Some magnetic systems, known as spin glasses, present random couplings which
lead to high levels of frustration and disordering. The metastability of
many structures of such spin glasses leads to experimental physical
measurements of the order of days due to slow decay of the magnetization to
zero. This peculiarity is also hard to explore in computer simulations.
Sherrington and Kirkpatrick \cite{Sherrington1975} proposed an interesting
exactly solvable spin-glass model that shows an slow dynamics of
magnetization.

Based on these annealing phenomena (heating treatment of materials to
enlarge its ductility) in solids, Kirkpatrick, Gelatt, and Vecchi \cite%
{Kirkpatrick1983} observed a deep connection between the Statistical
Mechanics of systems with many degrees of freedom, which can be governed by
the Metropolis prescription \cite{Metropolis1953} to thermal equilibrium,
and the combinatorial optimization of some problems related to obtain the
minimal of functions with many parameters in complex landscapes.

Many combinatorial optimization problems belong to the complexity class
known as NP-complete which is a subset of class simply known as NP class.
Let us better explain this fundamental point in the complexity of
algorithms. A problem belongs to the NP class if it can be decided in
polynomial time by a Turing machine (a universal computer model, for more
details see \cite{Papadmitriou}). Here it is important to say that not
necessarily it possesses an algorithm that can solve it in polynomial time.
Sure, we know that there are problems which can be solved by an algorithm in
polynomial time (fast solution) and not only decided! A simple example is
sorting a list.

One algorithm in these conditions belongs to a complexity class known as P,
and obviously P$\ \subset \ $NP. However there are some "rebel" problems in
NP, the so-called NP-complete problems, that can be understood, in a
non-rigorous way, as the harder problems in NP, since there is not a single
known algorithm to date, that can solve one of these problems in polynomial
time differently than occurs with problems in P.

Should we then believe in conjecture P$\ \neq \ $NP? Yes, and the reason is
simple! If we know one NP-complete problem, any algorithm that can solve it,
can be used to solve any other NP-complete problem (this is technically
called polynomial reduction, and for a good reference in complexity theory,
see again \cite{Papadmitriou}). Therefore, if somebody discovers an
algorithm in polynomial time that solves one of these NP-complete problems,
by consequence, all of them will also have solution in polynomial time.
However, such algorithms were never found, which makes stronger the
conjecture.

But what do you do with these problems? An alternative for them is to use
heuristics, and those which are physically motivated can be a good option.
Thus, based on Ref.~\cite{Kirkpatrick1983}, which proposed a metaheuristics
known as simulated annealing (SA), an algorithm which performs a random walk
in the configuration space until it reaches the global equilibrium or, at
worst, an interesting local minimum, the physicists, and computer scientists
have explored several different combinatorial problems, including for
example the study of thermodynamics of the protein folding \cite{Hansmann},
the evaluation of periodic orbits of multi-electron atomic systems \cite%
{Mauger}, and many others.

Basically, the SA algorithm works in the following way: we start with an
initial solution/configuration $\sigma _{old}$, and calculate the energy
(cost in a more general context) of this initial solution $E(\sigma_{old})$.
After that, another state is randomly chosen, which is denoted by $%
\sigma_{new}$, and this new solution is accepted with a probability that
follows the Metropolis prescription \cite{Metropolis1953}:

\begin{equation}
\Pr (\sigma _{old}\rightarrow \sigma _{new})=\min \left\{ 1,\exp \left[ - 
\frac{1}{T}\left( E(\sigma _{new})-E(\sigma _{old})\right) \right] \right\} ,
\end{equation}%
otherwise the system remains in the same $\sigma _{old}$ state. This process
is repeated until the ensemble sampled of the system reaches an equilibrium
at the given temperature. Finally, the temperature $T$ is decreased by a
cooling schedule. The temperature will be decreased until a state with low
enough energy is found.

Here, it is important to mention that many cooling schedules can be applied.
There are cooling schedules that asymptotically converge towards the global
minimum as the one that cools the system with a logarithm rule \cite%
{Geman1984}. However such schedule converges very slowly and requires a long
computation time.

There are good alternative schedules \cite{Nourani}, that although without a
rigorous guarantee of the convergence towards the global optimum, are
computationally faster. One of them is the geometric cooling schedule \cite%
{Laarhoven1987} which considers that at time $t$, the temperature is given
by $T=\alpha ^{t}T_{0}$, with $0<\alpha <1$. An interesting version of this
heuristic works with two loops: an internal and another external which can
be resumed by the Heuristic SAGCS (Simulated annealing with geometric
cooling schedule) resumed in the algorithm \ref{algorithm}.

\begin{algorithm}[H]
	\SetKwInOut{Parameters}{parameters}
	\SetAlgoLined
	
	\caption{SAGCS (Standard Simulated Annealing with Geometric Cooling Schedule)}
	
	\SetKwFunction{rand}{rand}
	\SetKwFunction{randomState}{randomState}
	
	\Parameters{
		$\sigma_0$ : initial configuration\\
		$T_{0}$ : initial temperature\\
		$T_{\min}$ : final temperature\\
		$N_{iter}$ : number of iterations of the internal loop\\
		$\alpha$ : coefficient of the geometric cooling
	}
	
	\KwResult{low energy state $\sigma$}
	
	\BlankLine
	
	$T      \gets T_0$\;
	$\sigma \gets \sigma_0$\;
	\While{$T > T_{\min}$}{
		\For{$i \gets 1$ \KwTo $N_{iter}$}{
			$\sigma_{new} \gets \randomState{}$\;
			\eIf{$E(\sigma_{new}) < E(\sigma)$}{
				$\sigma \gets \sigma_{new}$\;
			}{
				$x \gets \rand{$[0,1]$}$\;
				\If{$x < \exp{\left[-\frac{1}{T}(E(\sigma_{new})-E(\sigma))\right]}$}{
					$\sigma \gets \sigma_{new}$\;
				}
			}
		}
		$T \gets \alpha T$\;
	}
	\label{algorithm}
\end{algorithm}

\bigskip

It is important to observe that such Heuristic works with two parameters: $%
N_{iter}$ that controls the number of internal iterations, while the
external loop (iterations over different temperatures) is governed by:

\begin{equation}
N_{steps}=\left\lfloor \frac{\ln (T_{\min }/T_{0})}{\ln \alpha }\right\rfloor
\end{equation}%
and sure, for appropriate $N_{steps}$ and $N_{iter}$, the system must
converge for a local minimum that is very close to the global minimum.
Amongst the many interesting NP-complete problems, the traveling salesman
problem (TSP) deserves much attention and it became a natural way to test
the SA heuristic using different cooling schedules, variations of the
method, and parallelization techniques in many different contexts (see for
example \cite%
{Geman1984,Nourani,Laarhoven1987,Aarts1989,Azencott1992,Szu1987,Ingber1993}%
). The problem (TSP) is to find the optimal Hamiltonian cycle on a given
graph (vertices and valued edges), i.e., the minimum cost closed path
including all vertices passing through each vertex only once.

Let us consider graphs with $N$ vertices corresponding to points $%
(x_{i},y_{i})$, $i=1,..,N$, distributed in a two dimensional scenario. In
this case, the graph naturally has valued edges defined by the Euclidean
distances:

\begin{equation}
d(i,j)=\sqrt{(x_{i}-x_{j})^{2}+(y_{i}-y_{j})^{2}},  \label{Eq.dist}
\end{equation}
with $i,j=1,..,N$, and therefore, the simplest situation is to consider the
graph having all edges, since all costs $d(i,j)$ can be defined for all pair
of the points, which is denoted by a complete graph.

In this paper, we are not interested in testing cooling schedules, or even
other SA heuristics, which are very well explored in the literature, but in
building computer experiments to test the SAGCS in the TSP in order to
understand its efficiency considering the effects of correlation and
variance on the random coordinates $x_{i}$ and $y_{i}$. It is important to
mention that our method to include correlation is very simple and allows to
change the correlation but keeping the same variance which is very important
for making a fair comparison, since we simultaneously variate the two
parameters, we do not know if the effects are only caused by the correlation.

Our results show how the performance of the algorithm transits from
two-dimensional scenario ($\rho =0$) to the one-dimensional situation ($\rho
=1$) by deforming the region where the points (cities) are distributed. Here 
$\rho $\ is the correlation coefficient between the coordinates of the
points. We also investigate a parameter that controls the moments of the
random variables $x_{i}$ and $y_{i}$ and we quantitatively show how the
increase of the variance affects the performance. We particularly show that
the shape of the distribution is more important than its variance in certain
situations. On the other hand, when the points have coordinates that are
power-law distributed, in the region where the average cannot be defined,
the efficiency of the algorithm is negligible independently on scaling.

In the next section we will show some fundamental and pedagogical aspects of
the SA for the TSP. In the following, in section \ref%
{sec:correlated_and_long-tailed_environments} we define the different
environments where we will apply the SAGCS. We will show in detail how to
generate points with correlated coordinates, and how to generate points with
power-law distributed coordinates. In section \ref{sec:results} we show our
results and finally in section \ref{sec:conclusions} we present some
conclusions.

\section{Pedagogical aspects of the Simulated Annealing in the context of
the Travelling salesman problem}

\label{sec:SA}

The simulated annealing with geometric schedule (SAGCS) is a very simple
heuristic used to optimize combinatorial problems as the Traveling Salesman
Problem. The points $(x_{i},y_{i})$, with $i=1,...,N$ are scattered over a
two dimensional region, where the coordinates follow some joint probability
distribution $p(x_{i},y_{i})$.

Denoting by $v(i)\equiv (x_{i},y_{i})$ the $i$-th point of the cycle $\sigma
\equiv $ $v(1)\rightarrow v(2)\rightarrow ....\rightarrow v(N-1)\rightarrow
v(N)\rightarrow v(1)$, whose cost is given by 
\begin{equation}
C=E(\sigma )\sum_{k=1}^{N}d(v(k),v(k+1))
\end{equation}%
where $v(N+1)=v(1)$ and $d(i,j)$ is given by Eq. \ref{Eq.dist}. We aim to
determine the best cycle $\sigma $, i.e., the cycle of minimal cost.

The new configuration can be obtained performing different mechanisms in
this work. The first one is the \textquotedblleft simple
swap\textquotedblright (SS)\ , i.e., a point is randomly chosen, for example 
$v(j)$, and the candidate to be the new configuration $\sigma ^{\prime }\ $%
corresponds to $\sigma $ by simply changing the positions of $v(j)\equiv
(x_{j},y_{j})$ and $v(j+1)\equiv (x_{j+1},y_{j+1})$, then we can perform the
Metropolis prescription to optimize the problem in the context of the
simulated annealing.

On the other hand, there is a more interesting way to obtain a new
configuration according to \cite{Lin1973,KirkpatrickII} the 2-opt move. It
is a popular procedure used to improve algorithms to approximate the
solution of the Travelling Salesman Problem. Starting from a given cycle, it
consists in exchanging two links of the cycle to construct a new one. This
is performed by reversing the sequence of nodes between the selected links
and then reconnecting the cycle back together. For example, if one has a
route/cycle $\sigma \equiv v(1)\rightarrow v(2)\rightarrow v(3)\rightarrow
v(4)\rightarrow v(5)\rightarrow v(6)\rightarrow v(7)\rightarrow
v(8)\rightarrow v(1)$, by choosing $i=4$ and $i=7$, it leads to $\sigma
^{\prime }\equiv v(1)\rightarrow v(2)\rightarrow v(3)\rightarrow
v(7)\rightarrow v(6)\rightarrow v(5)\rightarrow v(4)\rightarrow
v(8)\rightarrow v(1)$. It allows us to explore the state space of
Hamiltonian cycles efficiently as well as it provides a simple way to
calculate the energy difference between the states before and after the
move, which is a very important feature for working in conjunction with the
Metropolis algorithm.

Now, let us explore the complexity of the problem. There are $N!$ different
cycles. Testing all of them is impossible for large $N$! For example for $%
N=2048$ cities (this will be a number of cities of the most of our
examples), by using the Stirling's formula $N!\approx \sqrt{2\pi }%
N^{N+1}e^{-N}$, one estimates 
\begin{equation*}
2048!\approx \sqrt{2\pi }2048^{2048+1}e^{-2048}\approx 7.57\times 10^{5895}
\end{equation*}

This a staggering number when compared, for example, with the number of
atoms of the universe $O(10^{80}$). Actually, among this "astronomical"
number of different cycles (much more than astronomical), certainly there
are a lot of them corresponding to a good local minima. It is exactly here
that SA shows its usefulness. Let us understand how it works by showing the
power of this algorithm. For example, let us examine the Fig. \ref%
{Fig:Importance_of_SA}.

\begin{figure}[t]
\begin{center}
\includegraphics[width=1.0		%
\columnwidth]{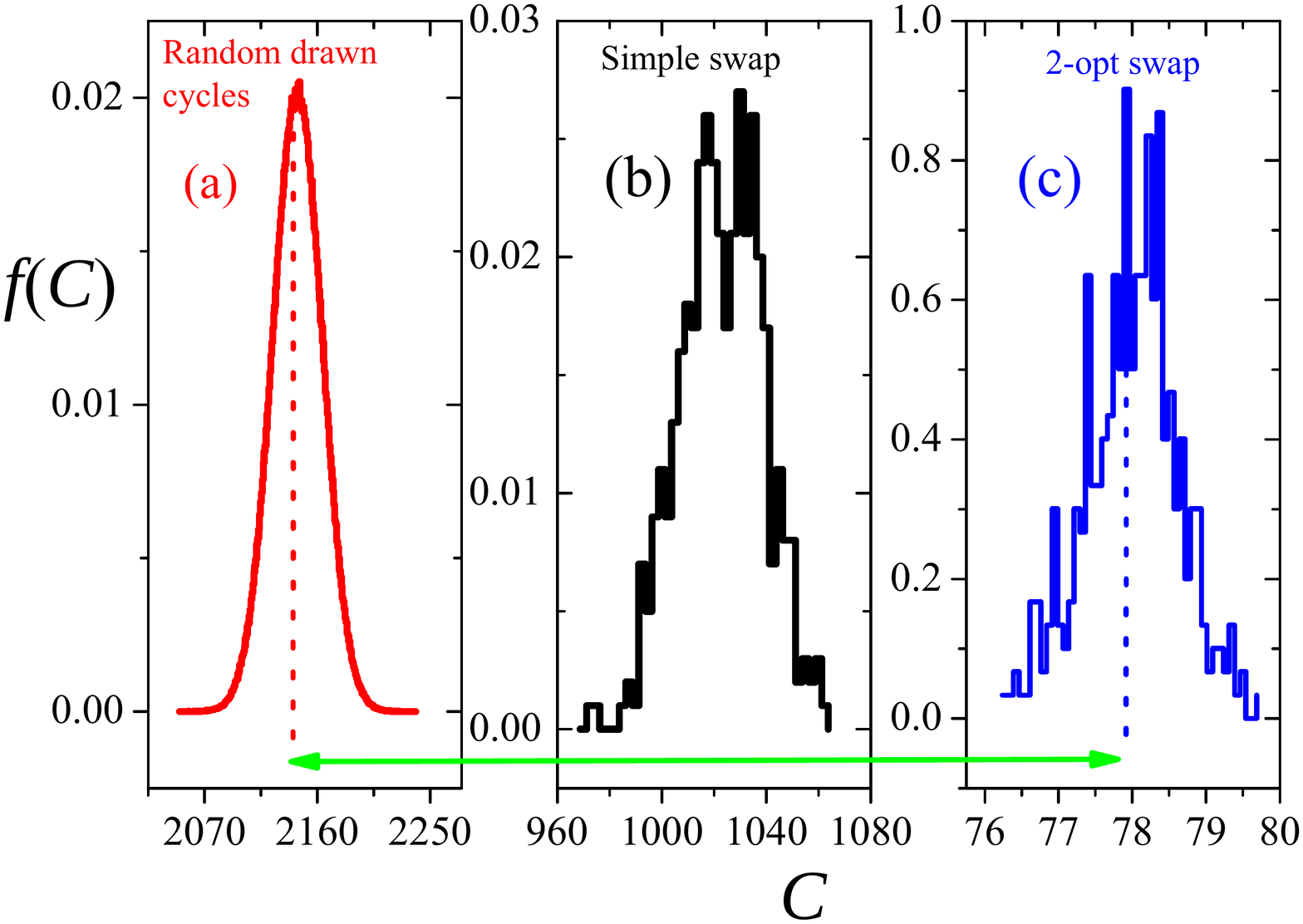}
\end{center}
\caption{{}The reason for that the SA is so important. We show the histogram
of the costs of the 10$^{6}$ random drawn cycles (a), 400 different runs of
the SA using simple swap (b), and 400 different runs of the SA using 2-opt
choice for new configurations (c). The line green is only to highlight the
huge difference between our best result obtained with SA and the
\textquotedblleft brute force sampling\textquotedblright }
\label{Fig:Importance_of_SA}
\end{figure}

This figure shows three histograms: the red one, represented in Fig. \ref%
{Fig:Importance_of_SA} (a) shows a sample of the costs of one million of
random drawn cycles of the same configuration with $N=2048$ cities, denoting
points whose coordinates are randomly distributed according to a uniform
distribution, with $x_{i}\in \lbrack -1,1]$ and $y_{i}\in \lbrack -1,1]$.
The black one, Fig. \ref{Fig:Importance_of_SA} (b), represents a histogram
with only 400 different costs obtained from $N_{run}=400$ different random
configurations of $N=2048$ of the SAGCS with simple swap (SS-SAGCS), by
considering $T_{0}=10$, $T_{\min }=10^{-5}$, $\alpha =0.99$, and $%
N_{iter}=8000$.

One can observe two very distinct Gaussian distributions whose means differ
across 1000 units of the cost. It is important to notice that not even one
among the one million costs of randomly drawn Hamiltonian cycles reached the
cost of SA. Actually, it is worse than this, in the first case we have $%
\overline{C}_{rand}=2143.74\pm 0.02$, while in the second case one has $%
\overline{C}_{SA}=1021.67\pm 0.74$ and the overlapping probability between
these two distributions is zero in the precision of the any known machine!
It is important to mention that we provocatively used only $N_{run}=400$
runs for the SA against the brute force (random drawn cycles).

The final blow comes when we use the 2-opt choice to sample new
configurations in the SAGCS (2opt-SAGCS), which is shown Fig. \ref%
{Fig:Importance_of_SA} (c), where we exactly used the same parameters of the
Fig. \ref{Fig:Importance_of_SA} (b). In this case one has $\overline{C}%
_{SA}=77.959\pm 0.031$ which is an extremely low cost when compared with
brute force method.

Thus, this pedagogical explanation is only to show that exhaustive sampling
is not a feasible solution for combinatorial problems. Heuristics as the SA
are an important alternative to obtain good (not always the optimal, but in
very disordered systems this is not be a meaningful difference) solutions to
the TSP. Now, after this preparatory study, we present the details about the
scenario for which we intend to analyze the performance of the SAGCS. In the
next section we will present the method to generate points with correlated
coordinates, and how to generate points with coordinates long-tailed
distributed. In these environments, we intend to explore some effects on the
performance of the SA which will be performed in section \ref{sec:results}.
\ 

\section{Correlated and long-tailed environments}

\label{sec:correlated_and_long-tailed_environments}

One of the important questions addressed in this paper is how the SAGCS
works considering that $x_{i}$ and $y_{i}$, the coordinates of the cities,
are random variables with a given correlation $\rho $. Additionally, for the
particular case $p(x_{i},y_{i})=p(x_{i})p(y_{i})$ ($\rho =0$), we also study
the performance of SAGCS for distributions $p(x_{i})\sim x_{i}^{-\gamma }$
and $p(y_{i})\sim y_{i}^{-\gamma }$, looking at the effect of $\gamma $ on
performance of the algorithm. Now, we discuss how to generate scenarios by
following these prescriptions.

\subsection{Generating correlated random coordinates from non-correlated
random variables}

In this section, we will show that we can generate correlated random
variables from non-correlated random variables considering that both
(correlated and non-correlated) have the same variance and average by
imposing an additional constraint -- considering the average equal to 0 for
the uncorrelated random variables. This is exactly the same procedure used
in \cite{rdasilva} in the context of emerging of rogue waves in the
superposition of electrical waves with correlated phases.

Let us consider spatial coordinates $x$ e $y$ of points (our cities) in a
two-dimensional environment that are random variables: 
\begin{equation}
\begin{array}{ccc}
x & = & \alpha _{1}z_{1}+\alpha _{2}z_{2} \\ 
&  &  \\ 
y & = & \beta _{1}z_{1}+\beta _{2}z_{2}%
\end{array}%
\end{equation}%
where $z_{1}$ and $z_{2}$ are independent and identically distributed random
variables, such that: $\left\langle z_{1}\right\rangle =\left\langle
z_{2}\right\rangle =\left\langle z\right\rangle $, and $\left\langle
z_{1}z_{2}\right\rangle =\left\langle z_{1}\right\rangle \left\langle
z_{2}\right\rangle =\left\langle z\right\rangle ^{2}$.

The variance of the variable $x$ is given by:$\ \left\langle \left( \Delta
x\right) ^{2}\right\rangle =\left\langle x^{2}\right\rangle -\left\langle
x\right\rangle ^{2}=(\alpha _{1}^{2}+\alpha _{2}^{2})\left\langle \left(
\Delta z\right) ^{2}\right\rangle $ where $\left\langle
z_{1}^{2}\right\rangle -\left\langle z_{1}\right\rangle ^{2}=\left\langle
z_{2}^{2}\right\rangle -\left\langle z_{2}\right\rangle ^{2}=\left\langle
z^{2}\right\rangle -\left\langle z\right\rangle ^{2}=\left\langle \left(
\Delta z\right) ^{2}\right\rangle $, and similarly $\left\langle \left(
\Delta y\right) ^{2}\right\rangle =(\beta _{1}^{2}+\beta
_{2}^{2})\left\langle \left( \Delta z\right) ^{2}\right\rangle $.

Now, we impose the condition 
\begin{equation}
\left\langle \left( \Delta x\right) ^{2}\right\rangle =\left\langle \left(
\Delta y\right) ^{2}\right\rangle =\left\langle \left( \Delta z\right)
^{2}\right\rangle ,  \label{Eq:equality_of_dispersions}
\end{equation}%
which implies that $\alpha _{1}^{2}+\alpha _{2}^{2}=\beta _{1}^{2}+\beta
_{2}^{2}=1$.

It is worth noting that although $z_{1}$ and $z_{2}$ are non-correlated
random variables, $x$ and $y$ are, and the correlation between these random
variables can be calculated: 
\begin{equation}
\rho =\frac{\left\langle \left( \ x-\left\langle x\right\rangle \right)
\left( \ y-\left\langle y\right\rangle \right) \right\rangle }{\sqrt{
\left\langle \left( \Delta x\right) ^{2}\right\rangle \left\langle \left(
\Delta y\right) ^{2}\right\rangle }}
\end{equation}

Thus, again after some cancellations and combinations: $\rho =(\alpha
_{1}\beta _{1}+\alpha _{2}\beta _{2})$. And after a little algebra, the
random variables can be now written as:%
\begin{equation}
x=\sin \left( \frac{1}{2}\sin ^{-1}(\rho )\right) z_{1}+\cos \left( \frac{1}{
2}\sin ^{-1}(\rho )\right) z_{2}  \label{Eq:var1}
\end{equation}%
and 
\begin{equation}
y=\cos \left( \frac{1}{2}\sin ^{-1}(\rho )\right) z_{1}+\sin \left( \frac{1}{
2}\sin ^{-1}(\rho )\right) z_{2}  \label{Eq:var2}
\end{equation}%
have the same average that are given by:

\begin{equation*}
\begin{array}{ccccc}
\left\langle x\right\rangle & = & \left\langle y\right\rangle & = & \frac{1}{
\sqrt{2}}\left[ \left( 1-\sqrt{1-\rho ^{2}}\right) ^{1/2}+\left( 1+\sqrt{
1-\rho ^{2}}\right) ^{1/2}\right] \left\langle z\right\rangle%
\end{array}%
\end{equation*}

From this, we can draw two important conclusions:

\begin{enumerate}
\item The random variables $x$ and $y$ have the same variance of $z_{1}$ and 
$z_{2}$ that are identically distributed and we required this according to
Eq. \ref{Eq:equality_of_dispersions} and therefore it does not depend on $%
\rho $;

\item If $\left\langle z_{1}\right\rangle =\left\langle z_{2}\right\rangle
=\left\langle z\right\rangle =0$, then $\left\langle x\right\rangle
=\left\langle y\right\rangle =0$.
\end{enumerate}

Thus, if one considers $x$ and $y$ as $\rho $-correlated random variables
generated from two independent random variables $z_{1}$ and $z_{2}$, with
average zero and variance $\sigma ^{2}=\left\langle \left( \Delta z\right)
^{2}\right\rangle $, $x$ and $y$ also have average zero and the same
variance $\sigma ^{2}$. This is an important point because if one works with
different averages and dispersions between the cases $\rho =0$ and $\rho
\neq 0 $\ the results can be misleading since there will be no parameter for
a fair comparison, i.e., in our problem no bias can be the source of the
possible studied effects, since independently on correlation, the random
variables are sampled with the same average and variance.

\begin{figure}[t]
\begin{center}
\includegraphics[width=1.0\columnwidth]{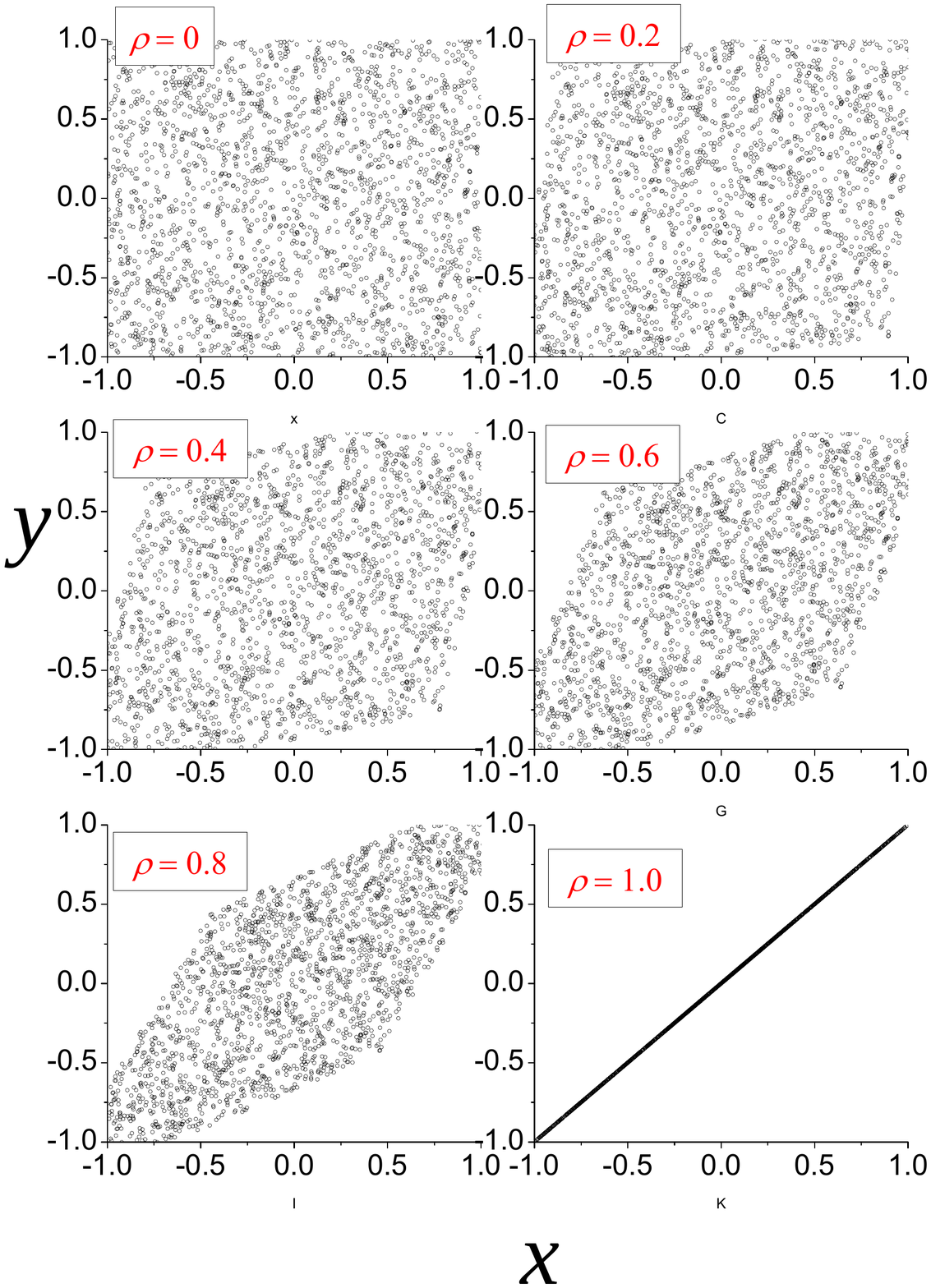}
\end{center}
\caption{{}Effects of correlation on the points using Eqs. \protect\ref%
{Eq:var1} and \protect\ref{Eq:var2}. }
\label{Fig:correlations}
\end{figure}
\qquad \qquad\ 

For example, using two uniform and identically distributed random variables $%
z_{1}=2(\xi _{1}-\frac{1}{2})$ and $z_{2}=2(\xi _{2}-\frac{1}{2})$ that
assume values in $[-1,1]$ generated from two $\xi _{1}$ and $\xi _{2}$
uniform random variables assuming values in $[0,1]$, the Fig. \ref%
{Fig:correlations} shows a plot of $y$ versus $x$ obtained from equations %
\ref{Eq:var1} and \ref{Eq:var2} for different values of $\rho $. In this
paper, the idea is to study the SAGCS on these correlated scenarios by
analyzing its performance.

\subsection{Power-law distributions for the coordinates}

Another important point of our study is to look at the effects of long
tailed distributions for the coordinates of the points on the SA
performance. Thus, we use a power-law probability density function to
generate the coordinates of the two-dimensional points. For that, we
initially propose the following distribution for the coordinates:

\begin{equation}
p(x;x_{0},\gamma )=\left\{ 
\begin{array}{cc}
\frac{(\gamma -1)}{2x_{0}^{1-\gamma }}\left\vert x\right\vert ^{-\gamma } & 
\text{if\ }\left\vert x\right\vert \geq x_{0} \\ 
&  \\ 
0 & \text{otherwise}%
\end{array}
\right.  \label{Eq:power_law}
\end{equation}

\begin{figure}[t]
\begin{center}
\includegraphics[width=1.0\columnwidth]{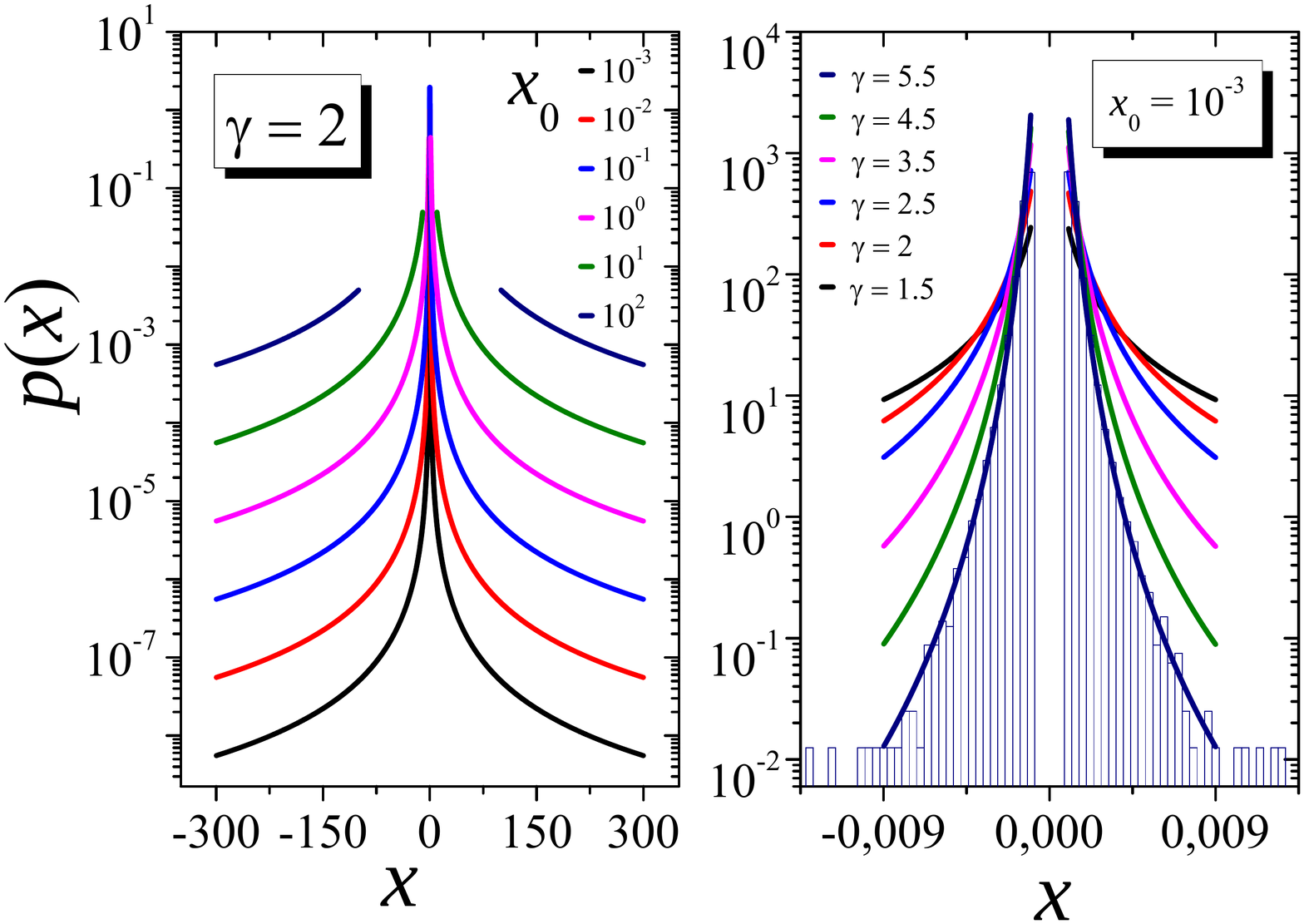}
\end{center}
\caption{Two-tailed power law distribution. Fig. (a): A plot of $p(x;x_{0}, 
\protect\gamma )$ for $\protect\gamma =2$ and different values of $x_{0}$.
Fig. (b): A plot of $p(x;x_{0},\protect\gamma )$ for $x_{0}=10^{-3}$ for
different values of $\protect\gamma $. The histogram is obtained with points
sampled with Eq. \protect\ref{Eq:x_long_range} for $\protect\gamma =5.5$,
only to check the formulae.}
\label{Fig:power_law_examples}
\end{figure}

It is important to observe that the power law distribution given by Eq. \ref%
{Eq:power_law} has a necessary gap $\Delta =2x_{0}$ with center in the
origin, since the two branches can not touch the origin by normalization,
however $x_{0}$ can be made arbitrarily small. Just for illustration, we
show plots of $p(x;x_{0},\gamma )$ in Fig. \ref{Fig:power_law_examples}.

\begin{figure}[t]
\begin{center}
\includegraphics[width=1.0\columnwidth]{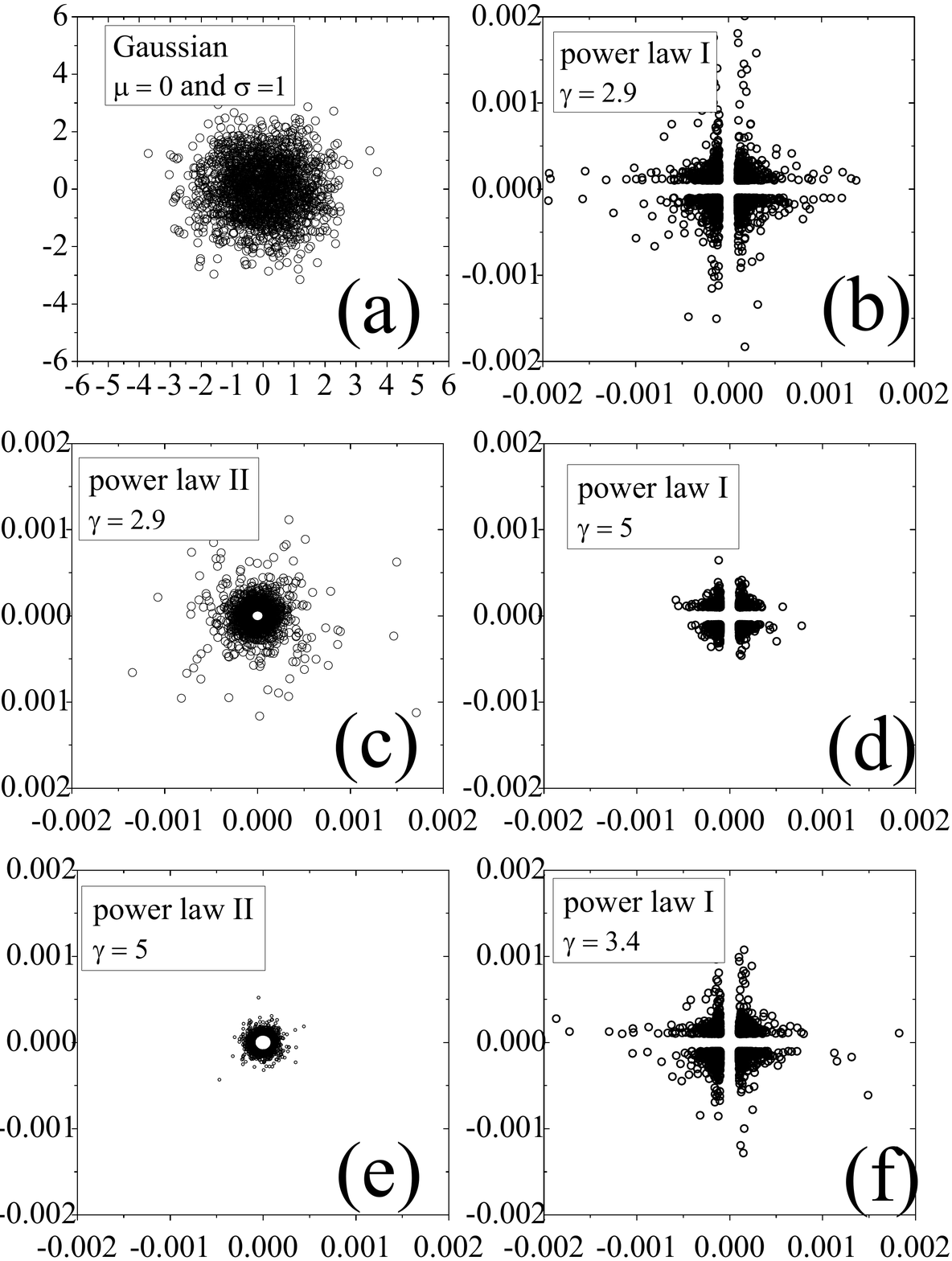}
\end{center}
\caption{Scattering of points for a comparison: (a) Gaussian coordinates,
(b-e) power-law coordinates for different approaches and for different
exponents $\protect\gamma $. It is important to notice that the figures are
shown with different intervals for a good visualization. We used both $x_{0}$
(power law I) and $r_{0}$ (power law II) equal to $10^{-3}$. }
\label{Fig:scattering_dists}
\end{figure}

In order to draw a variable that follow the distribution of Eq. \ref%
{Eq:power_law} one simply uses two uniform random (or more precisely
pseudo-random variables) variables $\xi _{1}$ and $\xi _{2}$ assuming values
in $[0,1]$. First, one chooses if the variable is on the positive or
negative branching. This is performed checking the sign of $(2\xi _{1}-1)$.
After that, one makes $\xi _{2}$ equal to area $\frac{(\gamma -1)}{%
x_{0}^{1-\gamma }}\int_{x_{0}}^{w}x^{-\gamma }dx$. This results in $%
w=x_{0}(1-\xi _{2})^{\frac{1}{1-\gamma }}$. Therefore, in the general case,
the random variable built from $\xi _{1}$ and $\xi _{2}$ is: 
\begin{equation}
x=\frac{2\xi _{1}-1}{\left\vert 2\xi _{1}-1\right\vert }x_{0}(1-\xi _{2})^{%
\frac{1}{1-\gamma }}  \label{Eq:x_long_range}
\end{equation}%
and naturally with other two random variables $\xi _{3}$ and $\xi _{4}$,
which are chosen from the uniform distribution in [0, 1], one has 
\begin{equation}
y=\frac{2\xi _{3}-1}{\left\vert 2\xi _{3}-1\right\vert }x_{0}(1-\xi _{4})^{%
\frac{1}{1-\gamma }}  \label{Eq:y_long_range}
\end{equation}%
In Fig. \ref{Fig:power_law_examples} (b), a histogram for the points $(x,y)$
sampled in this way is shown for $\gamma =5.5$, only to check the agreement
with the exact result. The idea is to observe the performance of the SAGCS
in this scenario answering how important is $\gamma $ and its effects on the
final cost obtained with SA algorithm. Notice that points generated by Eqs. %
\ref{Eq:x_long_range} and \ref{Eq:y_long_range} have border effects. For
instance, in Fig. \ref{Fig:scattering_dists}, we present some examples of
the points scattered according to different distributions by focusing the
power law ones.

Using the Gaussian distribution (with average zero and variance 1), the
pattern of points shown in Fig. \ref{Fig:scattering_dists} (a) is very
different than the case which uses the points generated by Eqs. \ref%
{Eq:x_long_range} and \ref{Eq:y_long_range} for $\gamma =2.9$ and $%
x_{0}=10^{-3}$ (Fig. \ref{Fig:scattering_dists} (b) ). Here, we will denote
the points using the Eqs. \ref{Eq:x_long_range} and \ref{Eq:y_long_range} by
power law I. We can observe some outlier points since, for this case, the
distribution has no second moment defined. However, radial symmetry is
absent in the power law I distribution and the border effects previously
mentioned are notorious. The question is whether they are really important.
Alternatively, in this paper, we also proposed a second power law type
(power law II) to study the SA, which respects the radial symmetry:

\begin{eqnarray}
x &=&r_{0}(1-\xi _{1})^{\frac{1}{1-\gamma }}\cos (2\pi \xi _{2})
\label{Eq:Power_law_II} \\
&&  \notag \\
y &=&r_{0}(1-\xi _{1})^{\frac{1}{1-\gamma }}\sin (2\pi \xi _{2})  \notag
\end{eqnarray}%
and in this case for $\gamma =2.9$ and $r_{0}=10^{-3}$ one has the
distribution of points represented in Fig. \ref{Fig:scattering_dists} (c).
Similarly $\xi _{1}$ and $\xi _{2}$ are uniform random variables assuming
values in $[0,1]$. Thus relevant questions should be related to the
performance of the SA in both situations: radially symmetric and asymmetric
ones. By completing in Fig \ref{Fig:scattering_dists} (d), and (e) we show
the distribution of points for a case where we have defined variance ($%
\gamma =5.5$) and therefore few outliers are observed. The case $\gamma =3.4$
complete our figure (Fig \ref{Fig:scattering_dists} (f) ) since this case
will be particularly important for our results. It is important to mention
that except by Fig. \ref{Fig:scattering_dists} (a) where the range used was $%
-4\leq x,y\leq 4$, all other Figs. \ref{Fig:scattering_dists} (b), (c), (d),
(e), and (f) that correspond to power-law cases, we used the range $%
-0.002\leq x,y\leq 0.002$\ for a suitable visualization and comparison.

\section{Results}

\label{sec:results}

We performed computer experiments to analyze two effects on the optimization
by the SA for the TSP: a) the correlation between the spatial coordinates,
and b) the variance/distribution shape for the spatial random coordinates.
We used the SAGCS which is a fast and standard way to perform optimization.
It is natural to expect that such effects must be proportionally important
in other variations of the SA employing other slower cooling schedules
independently if the final cost obtained is better. The goal of this paper
is not to compare different SA algorithms but performing a quantitative
study of the SAGCS considering different spatial distributions for the
coordinates of the points in the TSP.

However, before starting the main core of our results, it is important to
understand some preliminary aspects: the effects of the number of external
loop iterations $N_{steps}=\left\lfloor \frac{\ln (T_{\min }/T_{0})}{\ln
\alpha }\right\rfloor $, and $N_{iter}$ which denotes the number of internal
loop iterations on the final cost. We consider points $(x_{i},y_{i})$, $%
=1,...,N$ uniformly distributed non-correlated random variables (case $\rho
=0$ in Fig. \ref{Fig:correlations}). In all of our simulations, we averaged
our final cost over $N_{run}=60$ different runs of the SA.

So we performed simulations considering $T_{0}=10$ and $T_{\min }=10^{-5}$.
We initially consider the simple swap scheme to generate new candidate
configurations. Under these conditions, one obtains a plot of the cost in
the steady state ($C_{opt}$) which can be observed in Fig. \ref%
{Fig:cost_Ntotal} (a) as function of total number of iterations $%
N_{total}=N_{steps}N_{iter}$. Here we intent to show that such parameter is
more important than the parameters $N_{steps}(\alpha )$, and $N_{iter}$
separately. Our range for $N_{total}$ was from $5.10^{5}$ iterations up to $%
10^{9}$ iterations. The procedure is to change $N_{iter}$ from $2^{10}$ up
to $2^{18}$. Now, given the $N_{total}$ and the $N_{iter}$ we obtain the
factor of the exponent $\alpha $ of the cooling:%
\begin{equation*}
\alpha =\left\lfloor \left( \frac{T_{\min }}{T_{0}}\right) ^{\frac{N_{iter}}{%
N_{total}}}\right\rfloor
\end{equation*}

For example, when $N_{total}=5.10^{5}$, $\alpha $ changes from $0.001$ up to 
$0.9734$ and when $N_{total}=2.10^{9}$ one obtains that $\alpha $ changes
from $0.99831$ up to $0.99999$. There is actually a lot of information
aggregated in this figure to grasp. First the size of the points is
proportional to $N_{iter}$ and $\alpha \rightarrow 0$ corresponds to red,
while $\alpha \rightarrow 1$ corresponds to blue.

\begin{figure}[t]
\begin{center}
\includegraphics[width=1.0\columnwidth]{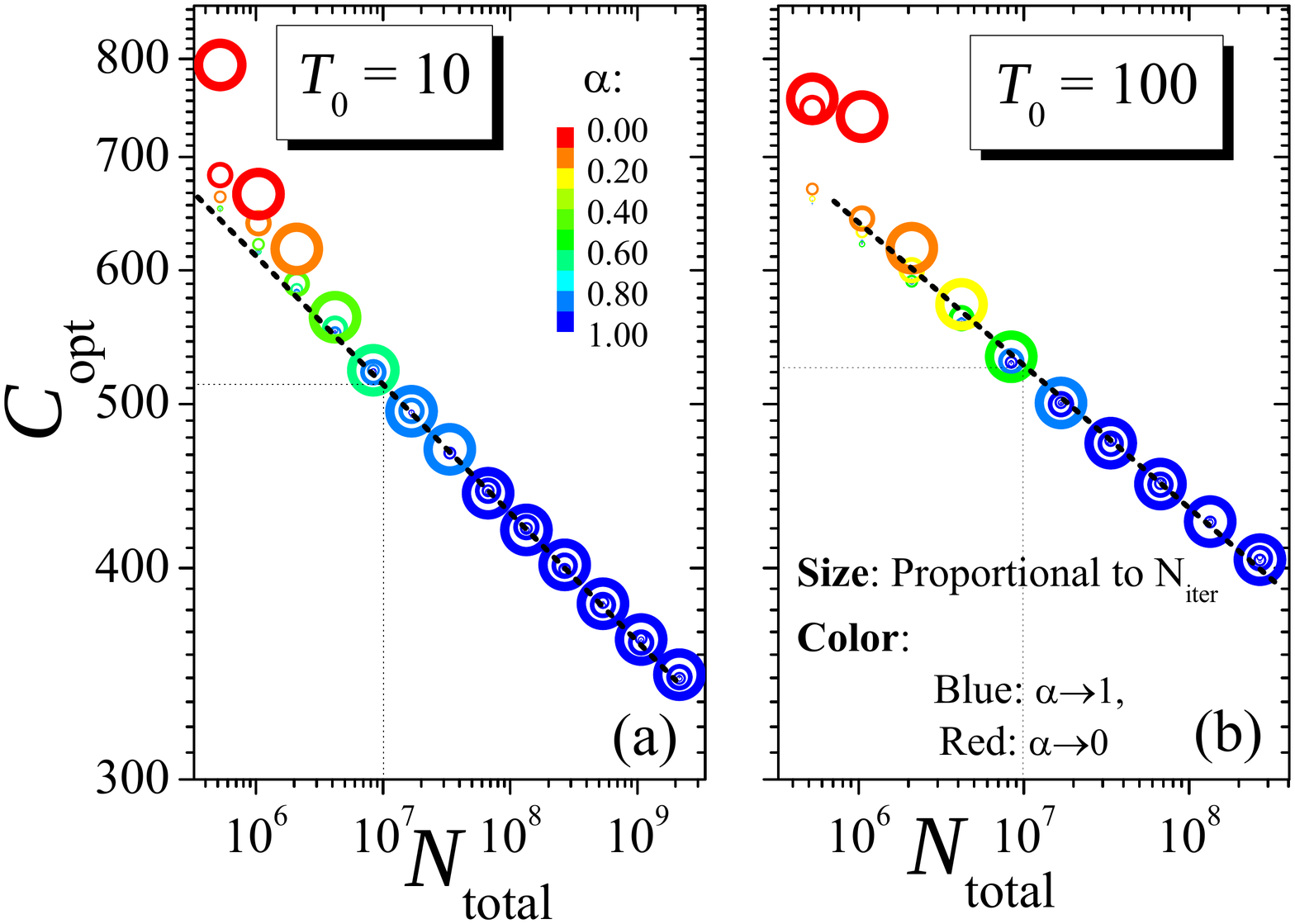}
\end{center}
\caption{{}Final cost (steady state cost) as function of $%
N_{total}=N_{steps}\cdot N_{iter}$. The color denotes the different values
of $\protect\alpha $ while the size of the points is proportional to $%
N_{iter}$. }
\label{Fig:cost_Ntotal}
\end{figure}
Fig. \ref{Fig:cost_Ntotal} shows that $C_{opt}$ depends on $N_{total}$ and
not on $N_{steps}$ and $N_{iter}$ separately. For this experiment it was
used $N=2048$ cities. The \ref{Fig:cost_Ntotal} (b) shows that for $%
T_{0}=100 $ the behavior is similar. The result suggests a reasonable
power-law universal behavior 
\begin{equation*}
C_{opt}\sim N_{total}^{-\delta }
\end{equation*}%
for sufficiently large total number of points in the sample independently on
the internal or external loops. For clarity, in the Fig. \ref%
{Fig:highlighting_power_laws}, we show a plot of $C_{opt}$ as function of $%
N_{total}$ for both $T_{0}=10$ and $T_{0}=100$ in the same plot, forgetting
at this moment the effects of $\alpha $ and $N_{iter}$ separately. In
log-log scale we obtain respectively $\delta =0.079\pm 0.001$ and $\delta
=0.083\pm 0.001$, which shows a similar power law behavior.

\begin{figure}[t]
\begin{center}
\includegraphics[width=1.0\columnwidth]{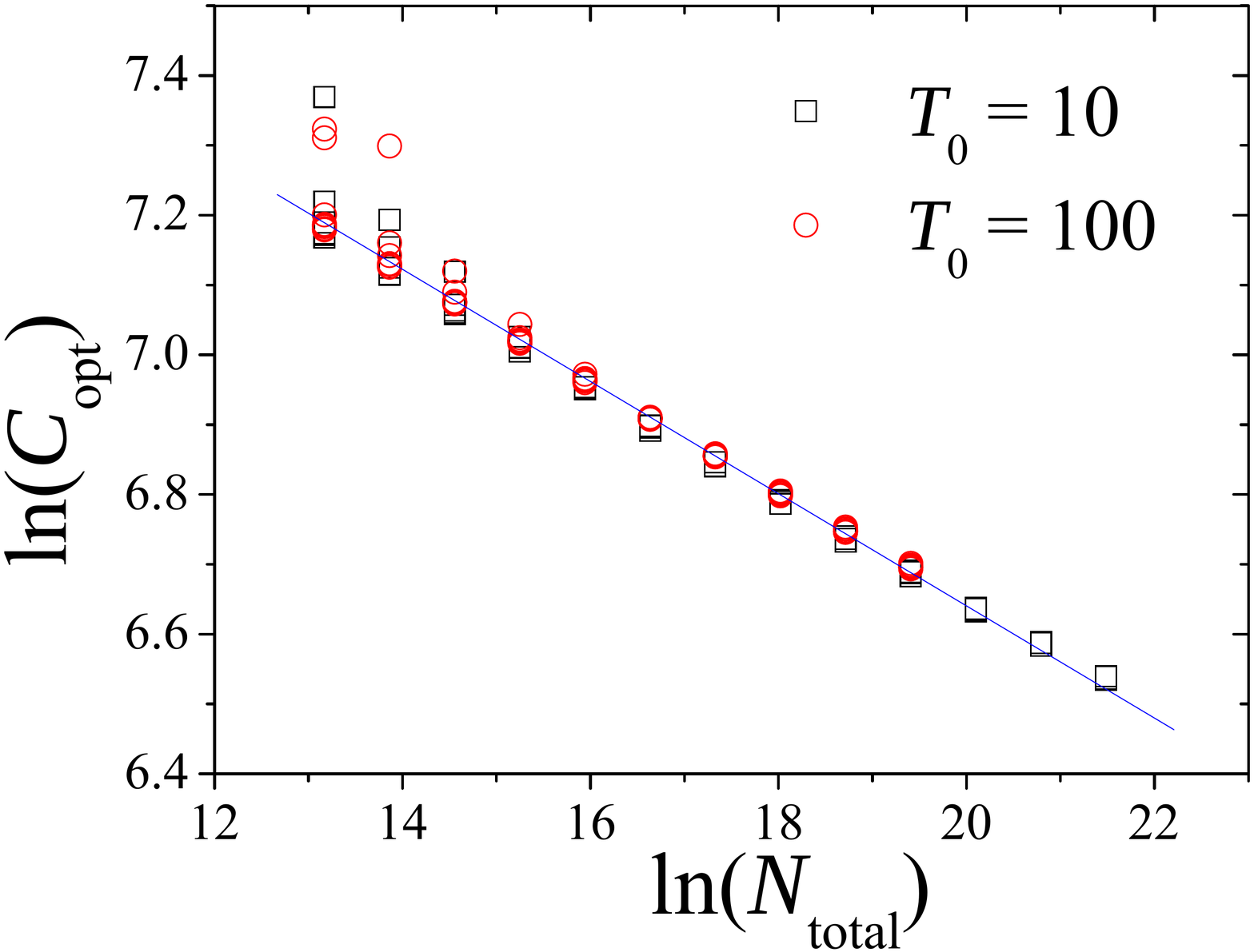}
\end{center}
\caption{{}Highlighting the power law behavior of the final cost as function
of $N_{total}$ discarding the effects of $N_{iter}$ and $\protect\alpha $
separately for both cases: $T_{0}=10$ and $T_{0}=100$. We can observe that
both curves have similar power law behaviors, where the obtained exponents
were $\protect\delta =0.079\pm 0.001$ and $\protect\delta =0.083\pm 0.001$
respectively. }
\label{Fig:highlighting_power_laws}
\end{figure}

\begin{figure}[t]
\begin{center}
\includegraphics[width=1.0\columnwidth]{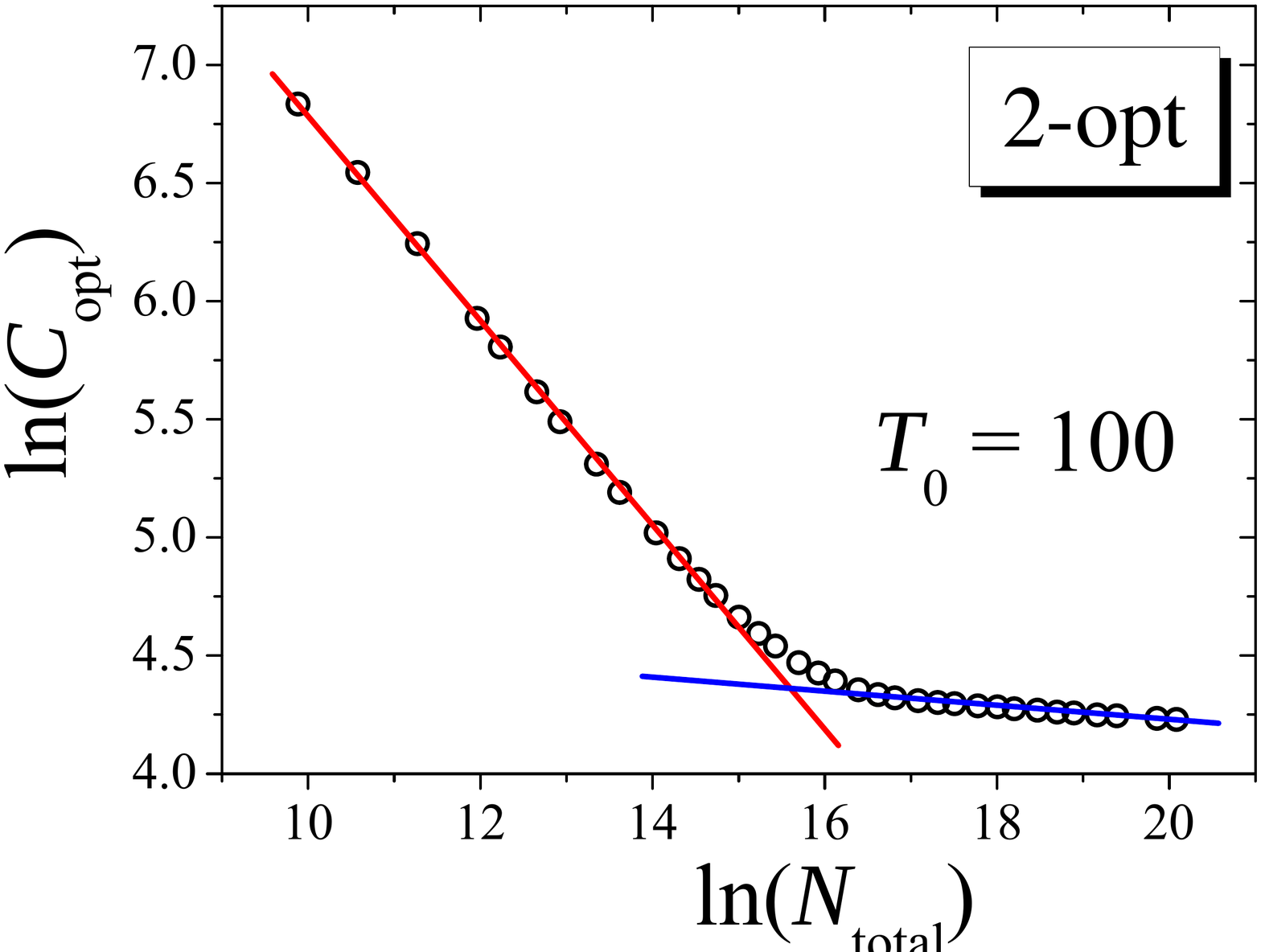}
\end{center}
\caption{{}Power law behavior of the final cost as function of $N_{total}$
for the prescription 2-opt exactly as performed in Fig. \protect\ref%
{Fig:highlighting_power_laws} for the simple swap. One observes a transition
from $\protect\delta \approx 0.43$ to $\protect\delta \approx 0.02$. }
\label{Fig:highlighting_2-opt}
\end{figure}

We perform a similar analysis of Fig. \ref{Fig:highlighting_power_laws} for
the 2-opt prescription, which is shown in Fig. \ref{Fig:highlighting_2-opt}.
In this case, somewhat differently from the simple swap, one observes a
transition between two power laws with very different exponents: from $%
\delta \approx 0.43$\ transiting to $\delta \approx 0.02$, which shows that
for $N_{total}>O(e^{16})\sim O(10^{6})$\ the optimal cost found does not
show a meaningful improvement, by changing from $C_{opt}\approx 90$\ up to $%
C_{opt}\approx 70$. After this preparatory study, we can proceed to the main
study of this work, to investigate the correlation and long range effects on
the coordinates of the points.

\subsection{Correlation effects}

From now, all of our results were obtained by using a fixed set of
parameters $N_{iter}=8000$, $\alpha =0.99$, $N_{run}=60$, $T_{\min }=10^{-5}$%
, and $T_{0}=10$, which results in $N_{total}=O(10^{7})$. Except by the
finite size scaling analysis, all of our experiments with SAGCS were applied
in scenarios with $N=2048$ cities.

\begin{figure}[t]
\begin{center}
\includegraphics[width=0.85\columnwidth]{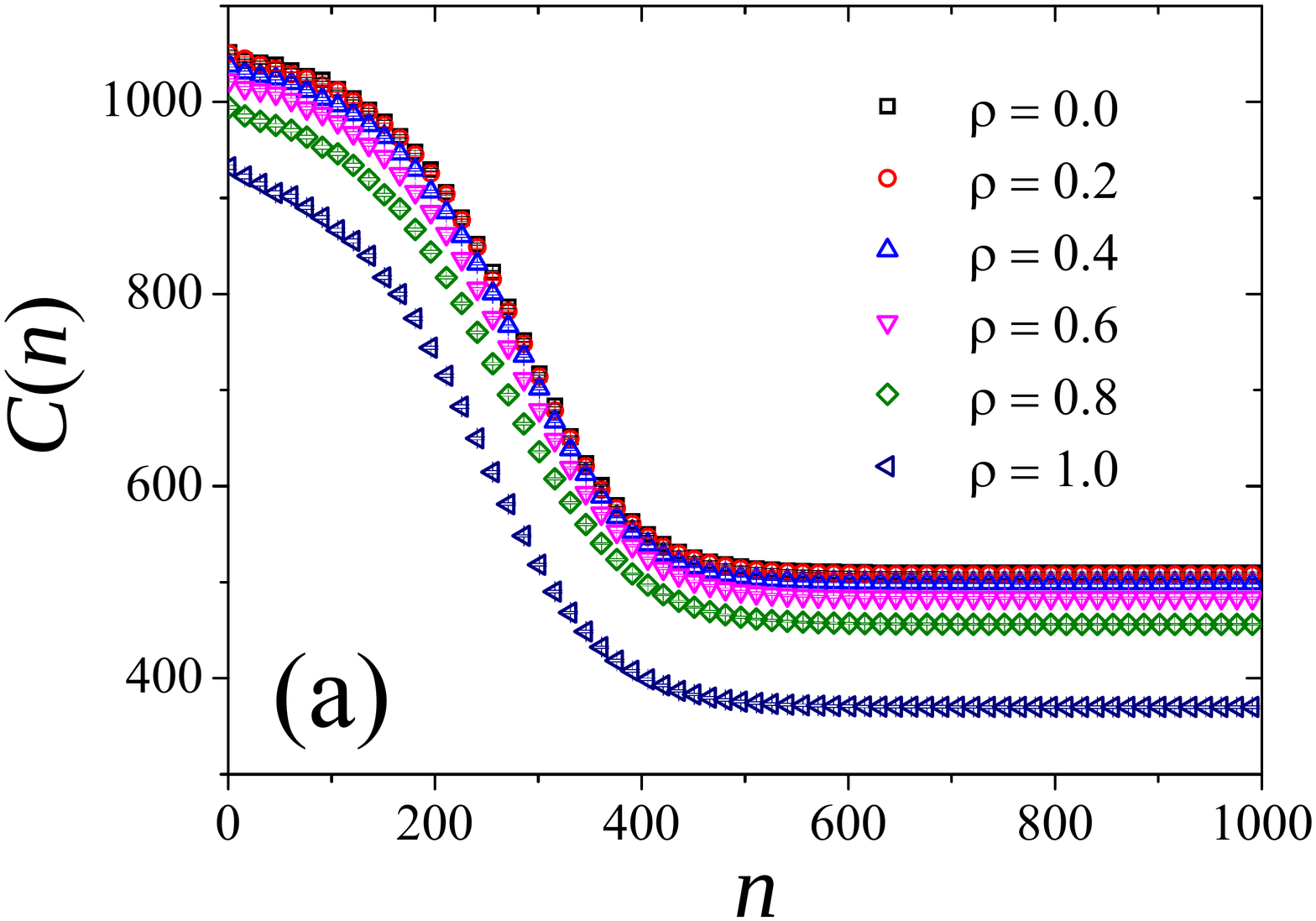} %
\includegraphics[width=0.85		%
\columnwidth]{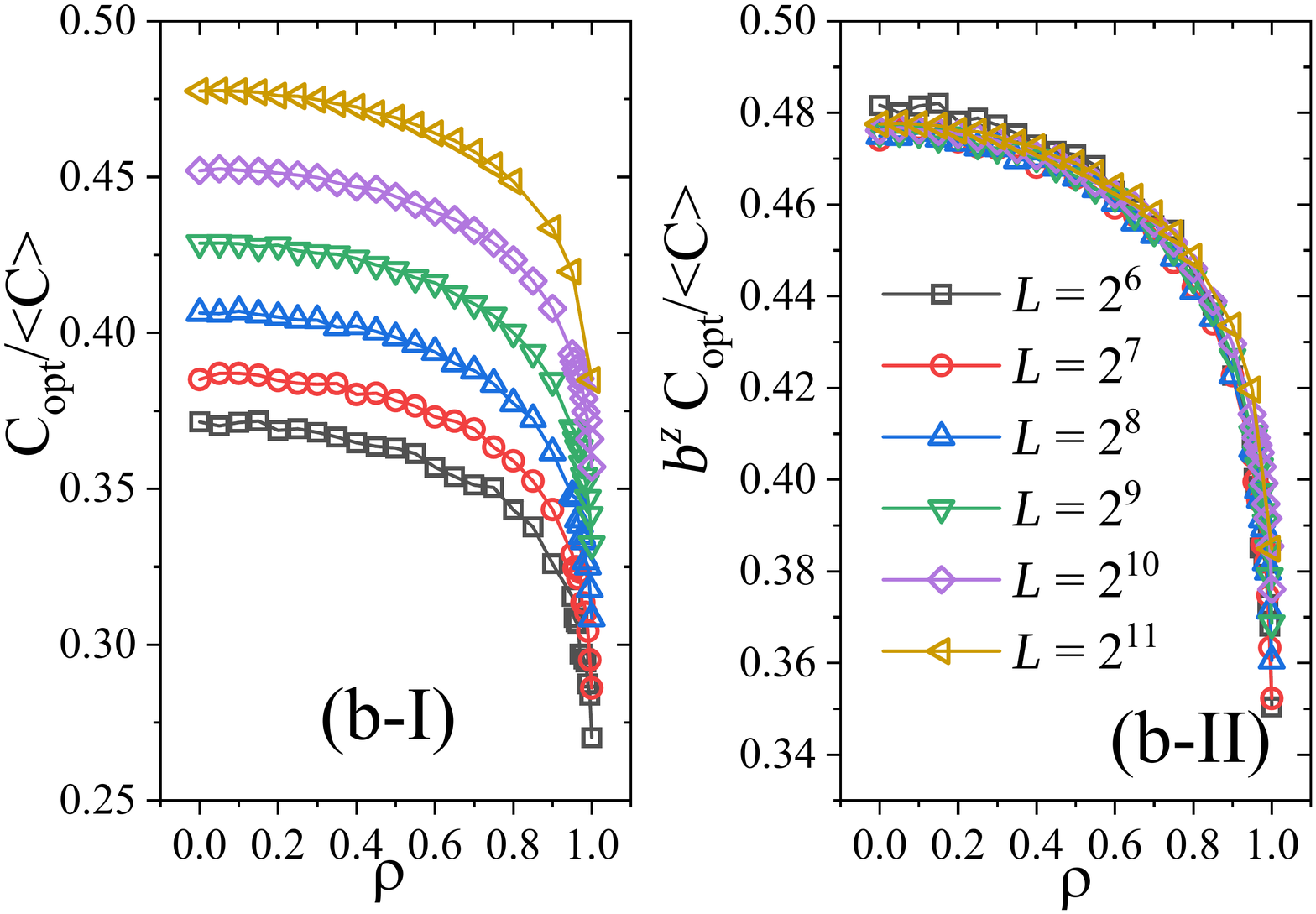}
\end{center}
\caption{Plot (a): Time evolution of the cost $C(n)$. Here $n$ is the
iteration of the external loop (cooling schedule) Plot b-I\textbf{:}
Performance of the SA as function of $\protect\rho $ for different number of
cities. Plot b-II An approximate scaling for the plot (b-I). One used
simple-swap prescription to generate new configurations in both plots.}
\label{Fig:cost_versus_rho}
\end{figure}

Let us start our study of correlation effects, by observing the behavior of $%
C$ as function of the iteration in the external loop (iteration of cooling
schedule) $n$, for different values of $\rho$ in scenarios with
cities/points ($x$,$y$) built from identical uniform random variables $z_{1}$
and $z_{2}$ defined on [-1,1] as the Fig. \ref{Fig:correlations}. We will
first study the SS-SAGCS performance.

In Fig. \ref{Fig:cost_versus_rho}(a) we can observe a decreasing of the
final cost $C_{opt}$ until it reaches a steady state after approximately 500
iterations. However, an interesting analysis is to consider the ratio $%
\overline{C_{opt}}/\overline{\left\langle C\right\rangle }$, given by 
\begin{equation}
\left\langle C\right\rangle =N\frac{\sum_{i<j}d(i,j)}{\binom{N}{2}}=\frac{2}{
(N-1)}\sum_{i<j}d(i,j),  \label{Eq:average_cost}
\end{equation}%
where the average cost is calculated considering that cities have an average
distance multiplied by the size of cycle, and here, $\overline{O}%
=(1/N_{run})\sum_{i=1}^{N_{run}}O_{i}$ denotes an average of amount $O$ over
the different runs (different steady costs obtained by the SA). It is
important to say that $\left\langle C\right\rangle $ calculated by Eq. \ref%
{Eq:average_cost} has no difference of the cost of any cycle which was
randomly chosen. This is a typical characteristic of the TSP, since
Hamiltonian cycles randomly generated are so far from the local minimum
found by heuristics as the SA, that even generating billions of such cycles,
the probability of reaching a good local minimum as the SA can be
disregarded as we previously discussed in Sec. \ref{sec:SA}.

The lower $C_{opt}/\left\langle C\right\rangle $, the better the
performance. Fig. \ref{Fig:cost_versus_rho} (b-I) shows the behavior of this
ratio as function of $\rho $ for several different values of $N$. We have an
improvement of the performance as $\rho $ enlarges. A reasonable (not so
good) empirical scaling can be observed in Fig. \ref{Fig:cost_versus_rho}%
(b-II). If one uses $b=L_{\max }/L$ and adjust $z\approx 0.075$, the
different curves have a good agreement in the collapse. Here $L_{\max }$\ is
the largest lattice size used in such study, which was $L_{\max }=2^{11}$%
\textbf{. }

We can observe that the performance for SS-SAGCS is even better for small
number of cities for the simple swap prescription. But, an important point
is to better investigate the performance of the SAGCS as function of $\rho $%
. Let us focus the extreme cases: $\rho =0$ and $\rho =1$. For $\rho =1$,
one has exactly the points scattered in a straight line (case $\rho =1$,
Fig. \ref{Fig:correlations}). In this particular case, we know in advance,
the global optimal Hamiltonian cycle, which is exactly given by $%
C_{glob}^{(theor)}=4\sqrt{2}$\ when $N\rightarrow \infty $, since we have
points increasingly closer to $(-1,-1)$\ and $(1,1)$\ the higher the value
of $N$, and therefore for finite $N$\ is only an approximation. On the other
hand, we can analytically estimate the averaged cycle 
\begin{equation*}
\begin{array}{lll}
\left\langle C\right\rangle _{\rho =1}^{(theor)} & = & \frac{2N}{N(N-1)}%
\sum_{i<j}^{N}d(i,j) \\ 
&  &  \\ 
& \approx & \frac{2N}{N(N-1)}\sum_{j=1}^{N-1}\sum_{i=1}^{N-j}i\left\langle
\Delta \right\rangle%
\end{array}%
\end{equation*}%
where $\left\langle \Delta \right\rangle $ $=\frac{4\sqrt{2}}{N}$ is the
average displacement between two adjacent pair of points which is randomly
distributed over the line. Thus, 
\begin{equation*}
\begin{array}{lll}
\left\langle C\right\rangle _{\rho =1}^{(theor)} & = & \frac{8\sqrt{2}}{%
N(N-1)}\sum_{j=1}^{N-1}\sum_{i=1}^{N-j}i \\ 
&  &  \\ 
& = & \frac{4\sqrt{2}}{3(N-1)}(N^{2}-1) \\ 
&  &  \\ 
& = & \frac{4\sqrt{2}}{3}(N+1)%
\end{array}%
\end{equation*}

Precisely: 
\begin{equation*}
C_{glob}^{(theor)}(\rho =1)/\left\langle C\right\rangle _{\rho
=1}^{(theor)}= \frac{3}{(N+1)}\text{.}
\end{equation*}%
which leads to the ratio $C_{glob}^{(theor)}/\left\langle C\right\rangle
_{\rho =1}^{(theor)}=\frac{3}{2049}\approx 0.001\,5$, for $N=2048$. This
shows that $C_{glob}$ is around of $O(10^{-3})$ smaller than $\left\langle
C\right\rangle _{\rho =1}^{(theor)}$, however the SA finds $\overline{C}%
_{opt}^{(SS)}/\overline{\left\langle C\right\rangle }\approx 0.38$ which is
a modest reduction. But one uses simple swap and such modest reduction is
indeed expected. What about 2-opt? With the same parameters, and using this
prescription, should the reduction of $O(10^{-3})$ be expected? Yes, as it
can be observed in Table \ref{Table:comparisom}, which shows a comparison of 
$C_{opt}$ for different values of $\rho $. One observes that $%
C_{opt}^{(2-opt)}$, the cost obtained with 2-opt-SAGCS is smaller than $%
C_{opt}^{(SS)}$ (the one obtained with SS-SAGCS) and the difference is huge
indeed.

Thus, considering the estimate for $\rho =1$, one obtains $\overline{C}%
_{opt}^{(2-opt)}/\overline{\left\langle C\right\rangle }\approx \allowbreak
0.002$ which leads to a very similar measure when compared with the one
obtained by analytical means.

\begin{table}[tbp] \centering
\begin{tabular}{|c|c|c|c|c|c|c|}
\hline\hline
$\rho $ & 0 & 0.2 & 0.4 & 0.6 & 0.8 & 1.0 \\ \hline\hline
Simple Swap: $\overline{C}_{opt}^{(SS)}$ & 1020.4(7) & 1014.5(7) & 997.9(7)
& 965.6(7) & 909.5(7) & 735.8(9) \\ \hline
2-Opt: $\overline{C}_{opt}^{(2-opt)}$ & 77.94(3) & 77.18(3) & 74.72(3) & 
69.90(3) & 60.82(2) & 7.873(4) \\ \hline\hline
\end{tabular}
\caption{Comparison between the simple swap and 2-opt prescriptions in the SAGCS for the final average costs 
		as function of $\rho$ }\label{Table:comparisom} 
\end{table}%

Actually, finding the optimal Hamiltonian cycle with points scattered in a
straight line is equivalent to order a list and one has good algorithms in
polynomial times (heapsort, quicksort...) that efficiently performs such
classification.

Nevertheless, it is important to mention that we have no previous
information about the topology of the points and the simulated annealing
with 2-opt prescription simply works to find a similar result to the optimal
one for $\rho =1$ as well as in any other situation, which is very
interesting indeed.

Let us analyze the ratio $C_{glob}/\left\langle C\right\rangle _{\rho =0}$
to check if the curve $C_{opt}/\left\langle C\right\rangle $ $\times \rho $
obtained with SA is compatible with these extremes. Once $C_{glob}$ for $%
\rho =0$ is not known, how do we estimate $C_{glob}/\left\langle
C\right\rangle _{\rho =0}$? First, let us start with a calculation that can
be performed exactly (see Appendix): 
\begin{equation*}
\left\langle C\right\rangle _{\rho =0}^{(theor)}=\frac{2N}{15}(2+\sqrt{2}
+5\ln (\sqrt{2}+1))
\end{equation*}%
which amounts to $2136$, approximately, for $N=2048$. Just to check, for $%
N_{run}=60$, we numerically obtain $\overline{\left\langle C\right\rangle }%
_{\rho =0}\approx2131$. Although we do not exactly know $C_{glob}$,
constructive heuristics (out of scope of the Boltzmann machines) can supply
an estimate which should work as a benchmark. A good suggestion is the
nearest neighbor (NN) algorithm \cite{Gutin2002}, a greedy algorithm that
corresponds to a particular case of the tourist random walk \cite%
{Martinez2001}, \cite{Stanley} with the maximal memory: $N-1$ can be an
candidate.

\begin{figure}[t]
\begin{center}
\includegraphics[width=0.85		%
\columnwidth]{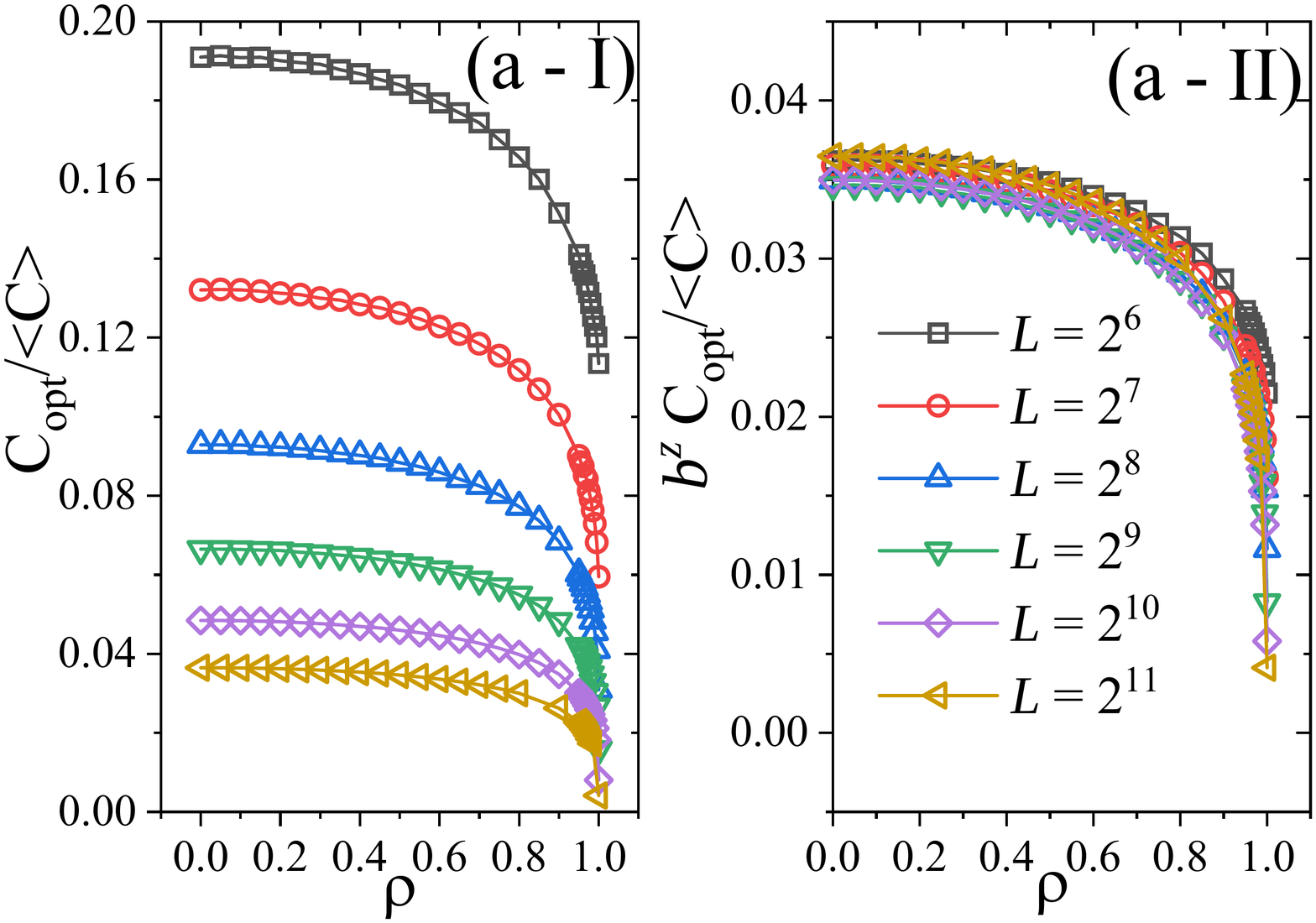} %
\includegraphics[width=0.85		%
\columnwidth]{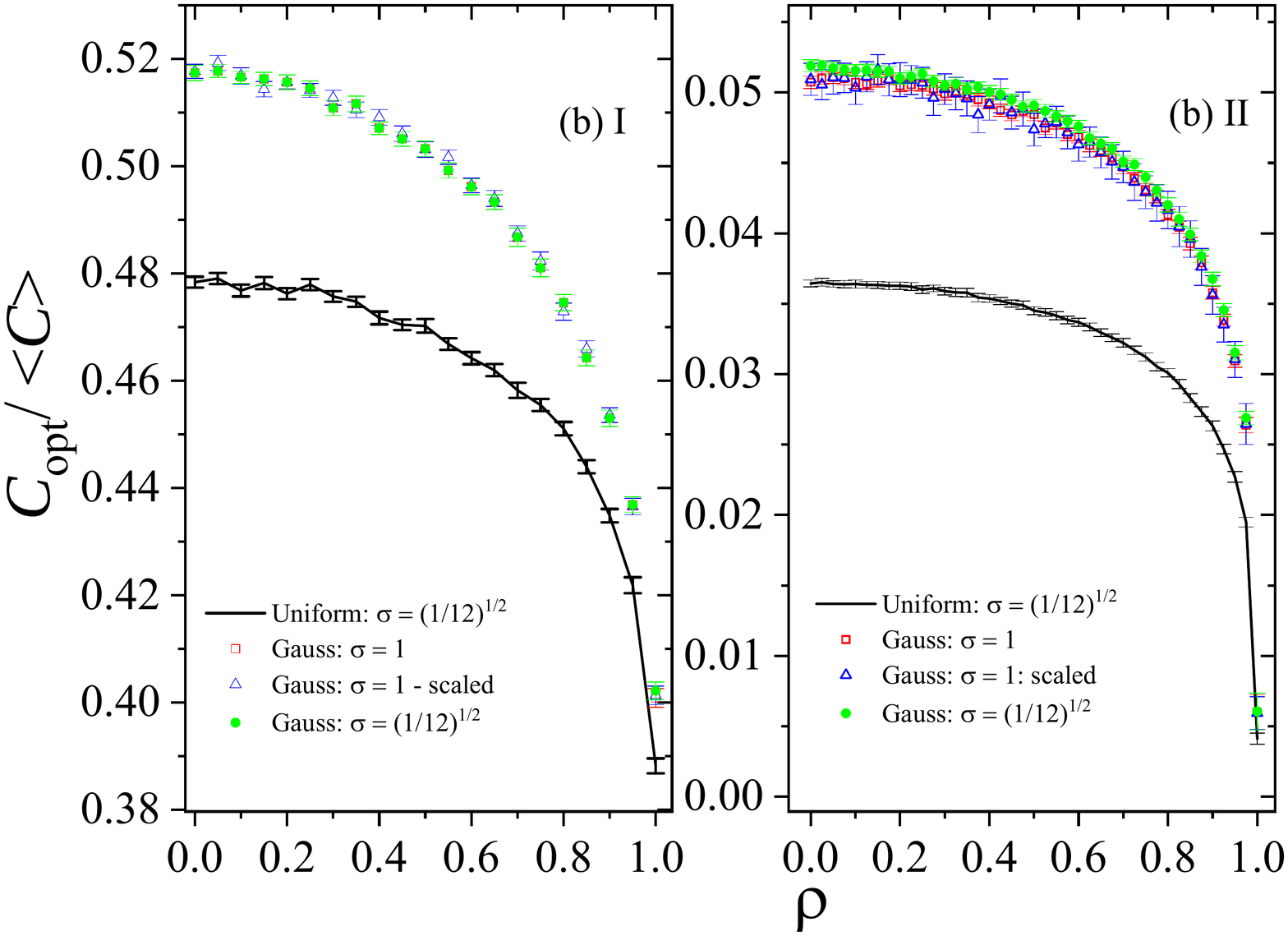}
\end{center}
\caption{(a) I: Performance of the 2-opt-SAGCS as function of $\protect\rho $
for different number of cities. (a) II: An approximate scaling for the plot
is obtained with $z\approx -0.480$. (b): Analysis for different
distributions of the points: Gaussian and uniform. The case (b) I
corresponds to SS-SAGCS and the (b) II corresponds to 2-opt-SAGCS.}
\label{Fig:scaling-2opt_and_dists_SS_and_2opt}
\end{figure}

In this algorithm the walk start from an initial node and jumps to the
nearest neighbor that has not yet been visited in the walk. The procedure is
repeated until returning to the initial node by closing the cycle
(Hamiltonian cycle). In this case, we applied this algorithm for $\rho =0$
and one finds a better approximation $\overline{C}%
_{opt}^{(NN)}(N=2048)=81.84\pm 0.34$ which leads to $\overline{C}%
_{opt}^{(NN)}/\overline{\left\langle C\right\rangle }_{\rho =0}\approx
\allowbreak 0.0384$ which is much better than $C_{opt}^{(SS)}/\overline{%
\left\langle C\right\rangle }_{\rho =0}\approx 0.479 $ which uses SS-SAGCS.
Surprisingly, the 2-opt-SAGCS $C_{opt}^{(2-Opt)}=77.94(3)$, which leads to $%
\overline{C}_{opt}^{(2-Opt)}/\overline{\left\langle C\right\rangle }_{\rho
=0}\approx 0.036\,$\ which is even better than the NN algorithm.

There exist many specially arranged city distributions which make the NN
algorithm gives the worst route. Here we are observing, it is an excellent
heuristic, however, the 2-opt-SAGCS is still a better benchmark (almost a
technical draw between the two heuristics) and the SAGCS has a good
complexity when compared with NN algorithm. The SA has a complexity that can
be writte in a general form as $O(c(N)\cdot N_{total})$\ where $c(N)=O(1)$\
for SS-SAGCS and $c(N)=O(N/2)$\ at the worst case for the 2-opt-SAGCS. On
the other hand the NN algorithm has a complexity $O(N^{2})$. Now it is
interesting to consider some points: $N_{total}$\ is not exactly related to $%
N$, since SA is a heuristic. Sure, $N_{steps}=\left\lfloor \frac{\ln
(T_{\min }/T_{0})}{\ln \alpha }\right\rfloor $, and therefore it is a
constant and does not depend on $N$. However, for larger values of $N$, it
is needed to make considerations about $N_{iter}$. It is natural to expect
that the larger $N$, larger the number of iterations of the internal loop
and we should consider $N_{iter}=O(N)$\ and the complexity of SA would be,
at the worst case, a complexity $O(N^{2})$\ which is exactly the complexity
of NN-algorithm considering the analysis of a single starting point. It is
important to mention that if we include the dependence on the initial point,
since different initial points can lead to different optimal costs, the
complexity of the NN-algorithm would then be $O(N^{3})$.

SA has modest optimal costs only when one uses a simple swap scheme, which
is a naive technique when compared with 2-opt that can applied in general
scenarios. In realistic situations, a generalization of the TSP, the VRP
(vehicle routing problem) is the best alternative compared to the NN
algorithm, and Tabu search algorithm as suggested by the authors in \cite%
{Wicaksono}.

But again, our proposal in this paper is to analyze possible effects of the
environment on the SA and not a detailed comparison among the methods, yet
we could not miss to show the 2-opt-SAGCS in comparison to SS-SAGCS and the
NN algorithm.

Thus, it is interesting to similarly analyze the size effects on the
2-opt-SAGCS exactly as we performed in plot in Fig. \ref{Fig:cost_versus_rho}
(b).

Fig. \ref{Fig:scaling-2opt_and_dists_SS_and_2opt} (a) I shows a similar
decay of $C_{opt}/\left\langle C\right\rangle $ as function of $\rho $ but
with values extremely lower than those of SS-SAGCS. Following what we
applied to the SS-SAGCS, we performed a similar scaling for the 2-opt-SAGCS
(Fig. \ref{Fig:scaling-2opt_and_dists_SS_and_2opt} (a) II) but not so with
the same quality. In this case the value is $z\approx -0.480$\ which is very
different than $z\approx 0.075$\ found to SS-SAGCS.

It is also interesting to analyze the effects on the points considering
different statistical distributions for the coordinates. Thus we prepared
some experiments to capture the effects on the ratio $C_{opt}/\left\langle
C\right\rangle $ as function of $\rho $ considering different distributions
for the coordinates and the result is also surprising as Fig. \ref%
{Fig:scaling-2opt_and_dists_SS_and_2opt} (b) shows. If before, we were
considering $z_{1}$ and $z_{2}$ identically distributed uniform random
variables, now we also consider them as Gaussian random variables for a
comparison. It is important to mention that the $x$ and $y$ coordinates,
have, the same average (zero) and variance of the variables $z_{1}$ and $%
z_{2}$ independently on $\rho $. As we previously reported, the effects to
be observed depends strictly on $\rho $ since the average and variance do
not change.

We can observe that in both cases, SS-SAGCS and 2-opt-SAGCS, respectively
described by Fig. \ref{Fig:scaling-2opt_and_dists_SS_and_2opt} (b-I) and
Fig. \ref{Fig:scaling-2opt_and_dists_SS_and_2opt} (b-II), the shape of the
distribution seems to be more important than the variance once the Gaussian
and the uniform random variables, with the same or different variances, lead
to different curves, but gaussians with different variances produce
practically the same behavior.

Finally, we also use an scaled cost: 
\begin{equation}
C_{s}=\frac{C}{\sqrt{(x_{\max }-x_{\min })(y_{\max }-y_{\min })}}\text{,}
\label{Eq:Scaling}
\end{equation}%
where $x(y)_{\max (\min )}=\max (\min )\{x(y)_{1},x(y)_{2},...,x(y)_{N}\}$,
once, differently from uniform distribution, the Gaussian distribution is
not supported on a finite interval. However, as the same Figs. \ref%
{Fig:scaling-2opt_and_dists_SS_and_2opt} (b-I) and (b-II) show, no changes
were observed in the two versions of the simulated annealing (the blue
triangles correspond to Gaussian distribution with scaled cost).

For SS-SAGCS the lower the $N$, the better the performance, on the other
hand, for the 2-opt-SAGCS, the higher the $N$, the better the performance as
also can be observed in the Figs. \ref{Fig:cost_versus_rho} (b-I) and \ref%
{Fig:scaling-2opt_and_dists_SS_and_2opt} (b-I).

\subsection{Finding good fits for $\overline{C}_{opt}/ \overline{%
\left\langle C\right\rangle }$ as a function of $\protect\rho $}

We also focus our results in finding good fits for $\overline{C}_{opt}/%
\overline{\left\langle C\right\rangle }$ as function of $\rho $. So follow a
(heuristic and qualitative) method to find suitable fits for such behavior.

\begin{figure}[t]
\begin{center}
\includegraphics[width=0.85\columnwidth]{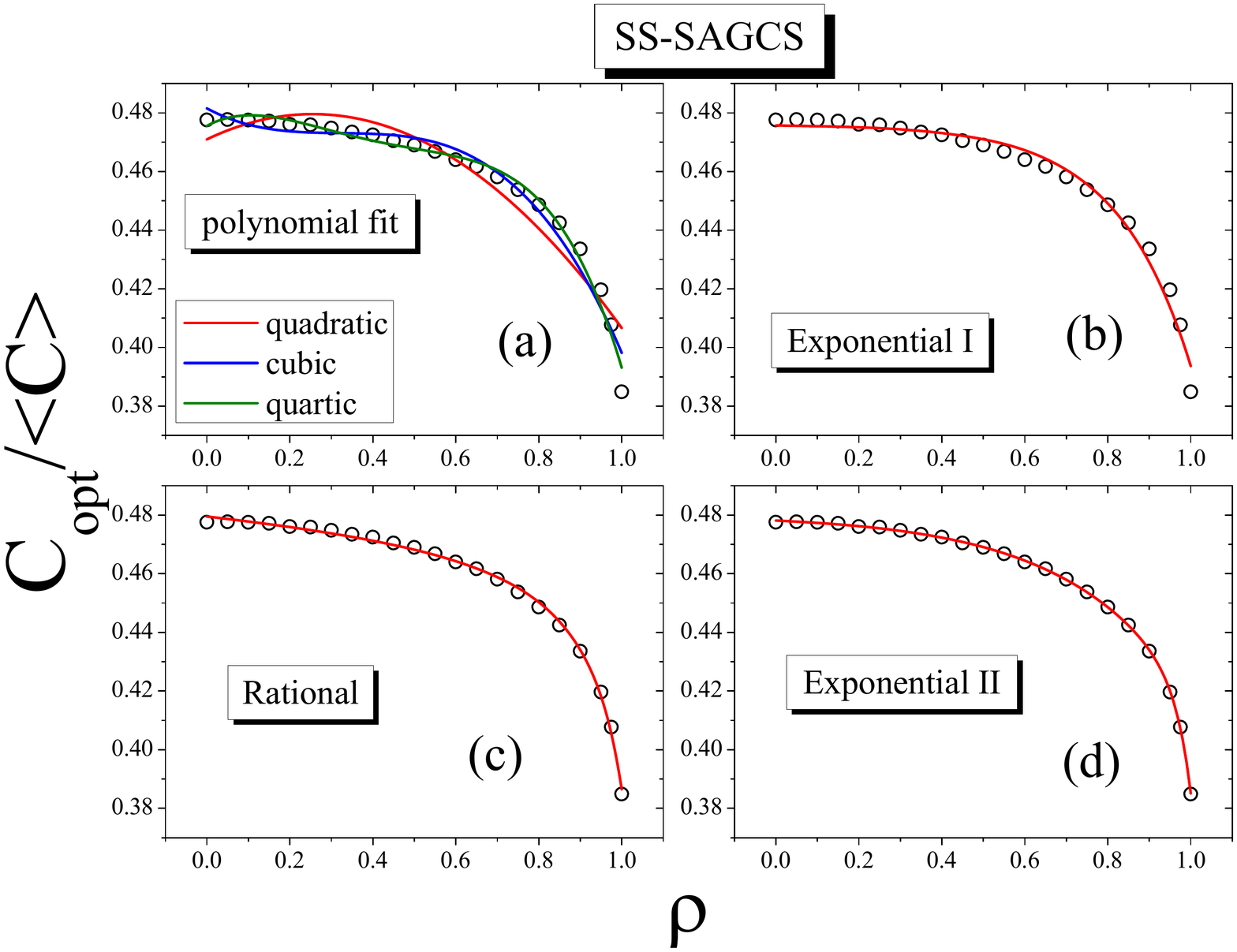}
\includegraphics[width=0.85\columnwidth]{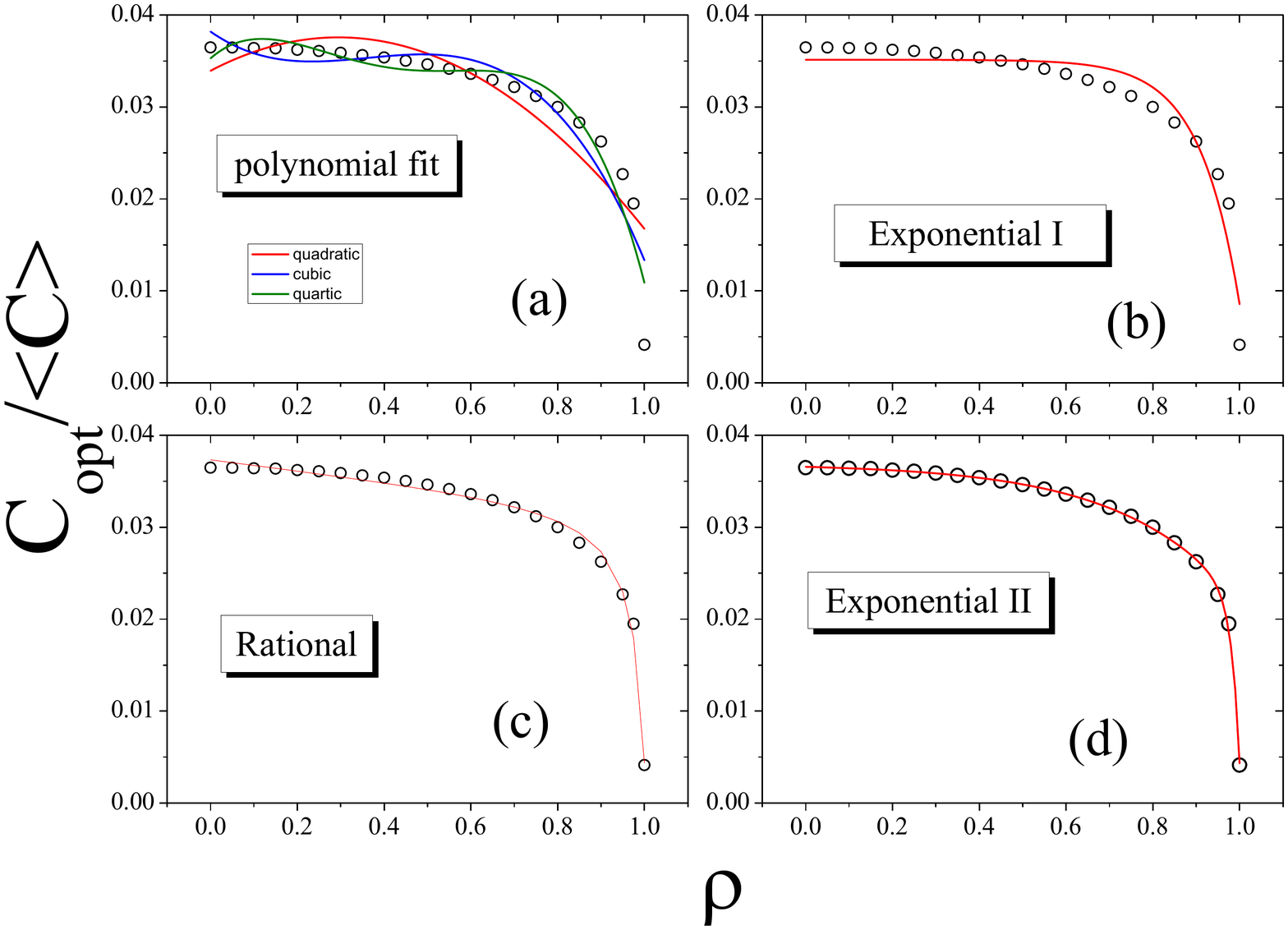}
\end{center}
\caption{Fits for performance versus $\protect\rho $ for SS-SAGCS and
2-opt-SAGCS. (a) preliminary polynomial fit (b) simple exponential (c) a
reasonable fit with rational function (d) the best fit obtained by combining
two exponentials.}
\label{Fits-SAGCS}
\end{figure}

First, we try a polynomial fit 
\begin{equation*}
p(\rho )=\sum_{i=0}^{n}a_{i}\rho ^{i}
\end{equation*}%
by testing $n=2$, $3$ and $4$ (with 3, 4, and 5 parameters respectively).
The idea here is only to make a preliminary investigation. After that, we
test exponential decay fits with only two parameters: 
\begin{equation*}
e_{I}(\rho )=e_{I}(0)+a_{1}e^{-\rho /\rho _{1}}
\end{equation*}%
once that $e_{I}(0)$ is fixed and taken as $\overline{C}_{opt}/\overline{%
\left\langle C\right\rangle }(0)$ (it was visually investigated). Following,
one considers a simple linear combination of exponential decays (with four
parameters):

\begin{equation}
e_{II}(\rho )=e_{II}(0)+a_{1}e^{-\rho /\rho _{1}}+a_{2}e^{-\rho /\rho _{2}} 
\text{,}  \label{Eq:linear_combination}
\end{equation}%
also assuming empirically that $e_{II}(0)=\overline{C}_{opt}/\overline{%
\left\langle C\right\rangle }(0)$.

Alternatively, we also experimented other functions with four parameters,
and the one that presented a good result was the rational function: 
\begin{equation*}
r(\rho )=\frac{a+b\rho }{1+c\rho +d\rho ^{2}}\text{.}
\end{equation*}

This is shown in Fig. \ref{Fits-SAGCS}, which shows the different fits. The
fits obtained in the different cases are summarized in table \ref{Table:fits}%
.

We can observe that the larger the degree of the polynomial fitted, the
better the coefficient of determination $\alpha $ (the closer to one, the
better it is). Our analysis also shows that in both cases, a single
exponential (exponential I) is not enough to nicely fit the curve but with
two exponential (four parameters), an excellent fit with $\alpha \approx
0.999$ can be obtained. With the same four parameters, the rational function
gives a good result despite not so good as the linear combination of
exponentials. It is important to mention that the coefficient $a$ found is
exactly $\overline{C}_{opt}/\overline{\left\langle C\right\rangle }(0)$ in
both cases (SS and 2opt-SAGCS) and the fit should be performed with three
parameters. After all, one can observe that a better match is obtained by
considering of a linear combination of the exponentials described by the Eq. %
\ref{Eq:linear_combination}. It seems to be a universal fit for $\overline{C}%
_{opt}/\overline{\left\langle C\right\rangle }$ $\times \rho $ for different
values of $\rho $ for both SS-SAGCS and 2-opt-SAGCS.

\begin{table}[tbp] \centering
\begin{tabular}{|l|l|l|l|}
\hline\hline
SA & $\text{Quadratic}$ & $\text{Cubic}$ & $\text{Quartic}$ \\ \hline\hline
& \multicolumn{1}{|l|}{$a_{0}=0.4710(42)$} & $a_{0}=0.4815(34)$ & $%
a_{0}=0.4755(26)$ \\ 
& \multicolumn{1}{|l|}{$a_{1}=0.067(19)$} & $a_{1}=-0.076(30)$ & $%
a_{1}=0.077(38)$ \\ 
SS & \multicolumn{1}{|l|}{$a_{2}=-0.132(18)$} & $a_{2}=0.230(70)$ & $%
a_{2}=-0.49(16)$ \\ 
& \multicolumn{1}{|l|}{$\alpha =0.92124$} & $a_{3}=-0.237(45)$ & $%
a_{3}=0.90(24)$ \\ 
& \multicolumn{1}{|l|}{} & $\alpha =0.96729$ & $a_{4}=-0.57(12)$ \\ 
& \multicolumn{1}{|l|}{} &  & $\alpha =0.98495$ \\ \hline
& \multicolumn{1}{|l|}{$a_{0}=0.0339(21)$} & $a_{0}=0.0382(21)$ & $%
a_{0}=0.0353(20)$ \\ 
& \multicolumn{1}{|l|}{$a_{1}=0.0247(96)$} & $a_{1}=0.0330(19)$ & $%
a_{1}=0.041(29)$ \\ 
2--opt & \multicolumn{1}{|l|}{$a_{2}=-0.0419(91)$} & $a_{2}=0.104(43)$ & $%
a_{2}=-0.25(12)$ \\ 
& \multicolumn{1}{|l|}{$\alpha =0.78373$} & $a_{3}=-0.096(28)$ & $%
a_{3}=-0.46(19)$ \\ 
& \multicolumn{1}{|l|}{} & $\alpha =0.86162$ & $a_{4}=-0.275(92)$ \\ 
& \multicolumn{1}{|l|}{} &  & $\alpha =0.90423$ \\ \hline\hline
\end{tabular}

\begin{tabular}{|l|l|l|l|}
\hline\hline
SA & $\text{Exponential I}$ & Rational & Exponential II \\ \hline\hline
& $c_{1}=-3.1(1.2)\cdot 10^{-4}$ & $a=0.47951(65)$ & $c_{1}=-0.00190(13)$ \\ 
& $\rho _{1}=0.179(13)$ & $b=-0.4443(29)$ & $\rho _{1}=-0.2852(62)$ \\ 
SS & $\alpha =0.98309$ & $c=-0.8923(39)$ & $c_{2}=-(2.4\pm 3.3)\cdot
10^{-15} $ \\ 
&  & $d=-0.0165(51)$ & $\rho _{2}=-0.0331(15)$ \\ 
&  & $\alpha =0.99798$ & $\alpha =0.99984$ \\ 
&  &  &  \\ \hline
& $c_{1}=-4.9(7.9)\cdot 10^{-7}$ & $a=0.03733(34)$ & $c_{1}=-1.5(5.4)\cdot
10^{-26}$ \\ 
& $\rho _{1}=-0.092(14)$ & $b=-0.03719(35)$ & $\rho _{1}=-0.0180(12)$ \\ 
2--opt & $\alpha =0.93539$ & $c=-0.832(24)$ & $c_{2}=-3.27(61)\cdot 10^{-4}$
\\ 
&  & $d=-0.136(25)$ & $\rho _{2}=-0.261(14)$ \\ 
&  & $\alpha =0.99312$ & $\alpha =0.99914$ \\ 
&  &  &  \\ \hline\hline
\end{tabular}
\caption{Values of coefficients found after nonlinear fitting by using Levenberg-Marquadt method. The parameter 
		$\alpha$ corresponds to coefficient of determination of each fit}\label%
{Table:fits} 
\end{table}%

\subsection{Long tail effects}

Finally, it is important to analyze the effects of long tailed distributions
for the coordinates of the points distributed on the environment. In this
case, we concentrate our analysis on $\rho =0$. Firstly, one chooses the
variables $x$ and $y$ according to equations \ref{Eq:x_long_range} and \ref%
{Eq:y_long_range} for the SS-SAGCS. Plotting $C_{opt}/\left\langle
C\right\rangle $ as function of $\gamma $ according to Fig. \ref%
{Fig:long_range} (a) and (b).

\begin{figure}[t]
\begin{center}
\includegraphics[width=0.85\columnwidth]{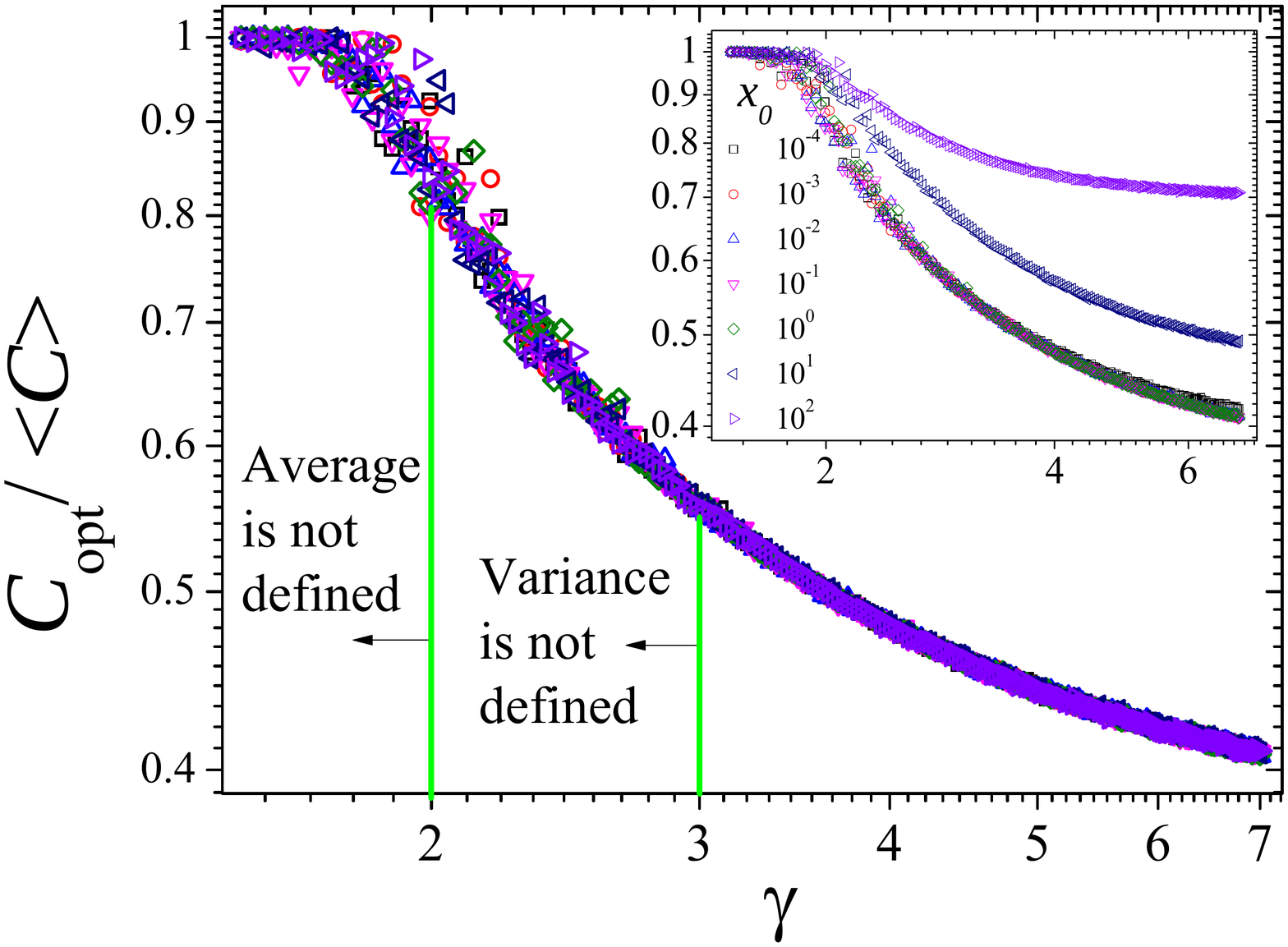} %
\includegraphics[width=0.85\columnwidth]{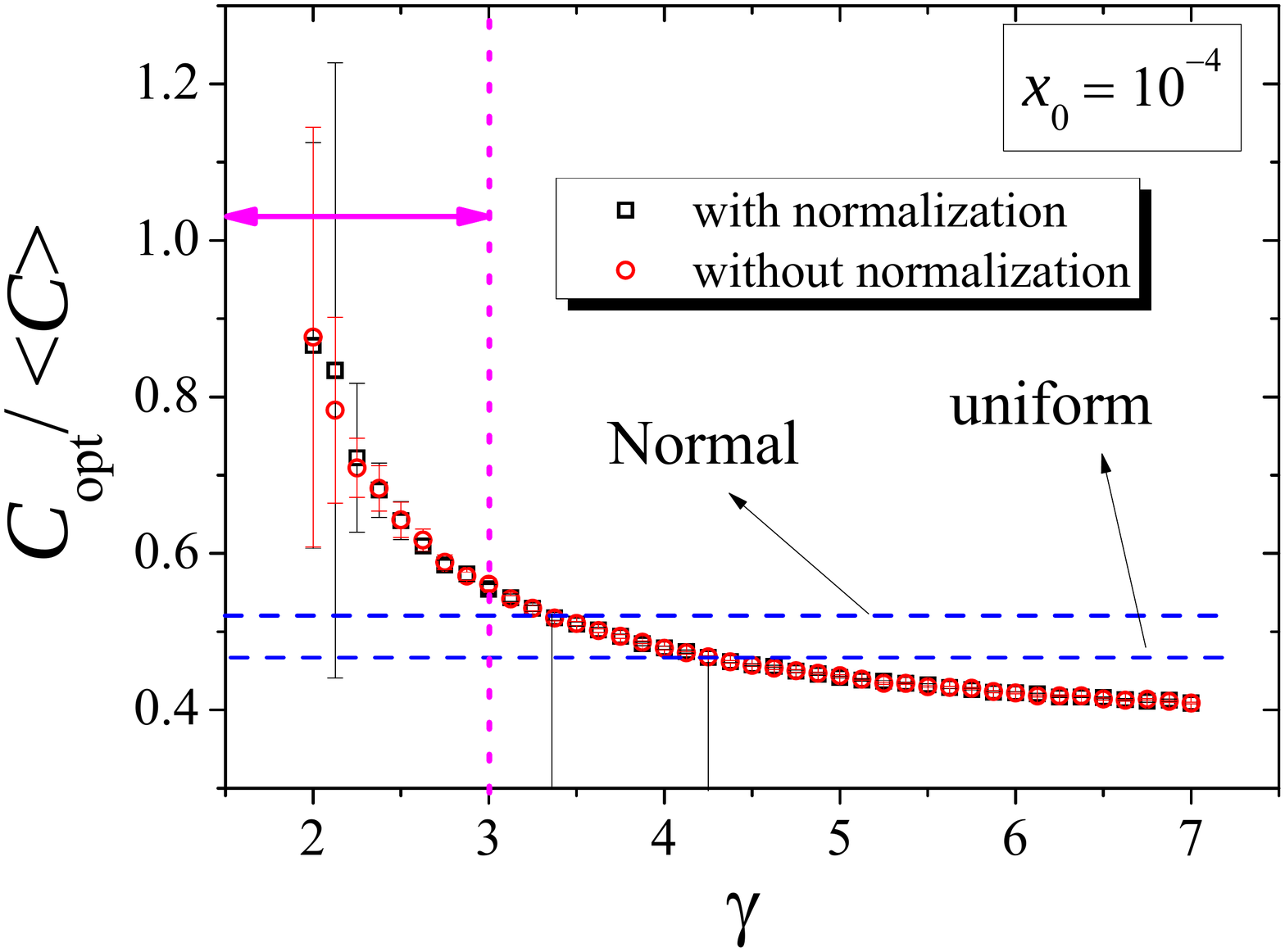}
\end{center}
\caption{\textbf{(a)} $C_{opt}/\left\langle C\right\rangle $ as function of $%
\protect\gamma $ for different values of $\protect\varepsilon $ for the
SS-SAGCS. We perform the scaling $C_{s}=C/\protect\sqrt{(x_{\max }-x_{\min
})(y_{\max }-y_{\min })}$. The inset plot shows the same curves without
scaling. \textbf{(b)} Refinement of the region presenting the error bars. We
can observe that $\protect\gamma \approx 3.4$ corresponds to Gaussian
distribution and $\protect\gamma \approx 4.3$ corresponds to uniform
distribution. }
\label{Fig:long_range}
\end{figure}

As expected $C_{opt}/\left\langle C\right\rangle $ decreases as function of
exponent $\gamma $. Fig. \ref{Fig:long_range} (a) shows this behavior for
different values of $x_{0}$ considering the scaling in Eq. \ref{Eq:Scaling}.
We can observe a collapse of all curves independently on $x_{0}$. The inset
plot shows the same plot without performing the scaling. It is interesting
to notice the instability in the regions without defined variance $2\leq
\gamma \leq 3$ but yet with some efficiency. For $\gamma <2$, which
corresponds the region where the average cannot be defined, the SA has no
efficiency which suggests that the algorithm should not be applied in these
situations since $C_{opt}/\left\langle C\right\rangle \approx 1$. Fig. \ref%
{Fig:long_range} (b) is only a refinement of one case in Fig. \ref%
{Fig:long_range} (a) with uncertainty bars and also showing the values
obtained with Gaussian and uniform distributions for a comparison (dashed
blue lines).

\begin{figure}[t]
\begin{center}
\includegraphics[width=0.85\columnwidth]{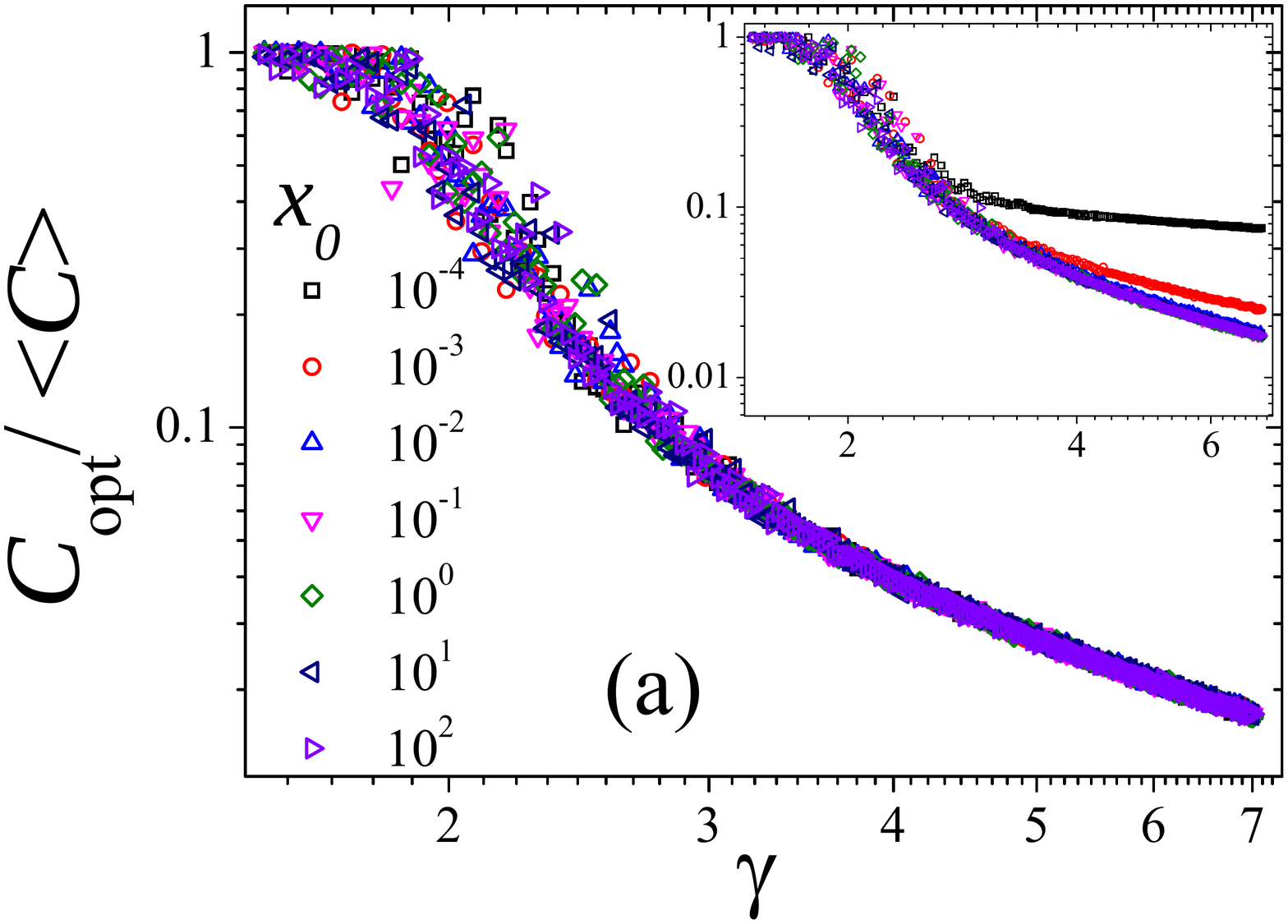} %
\includegraphics[width=0.85\columnwidth]{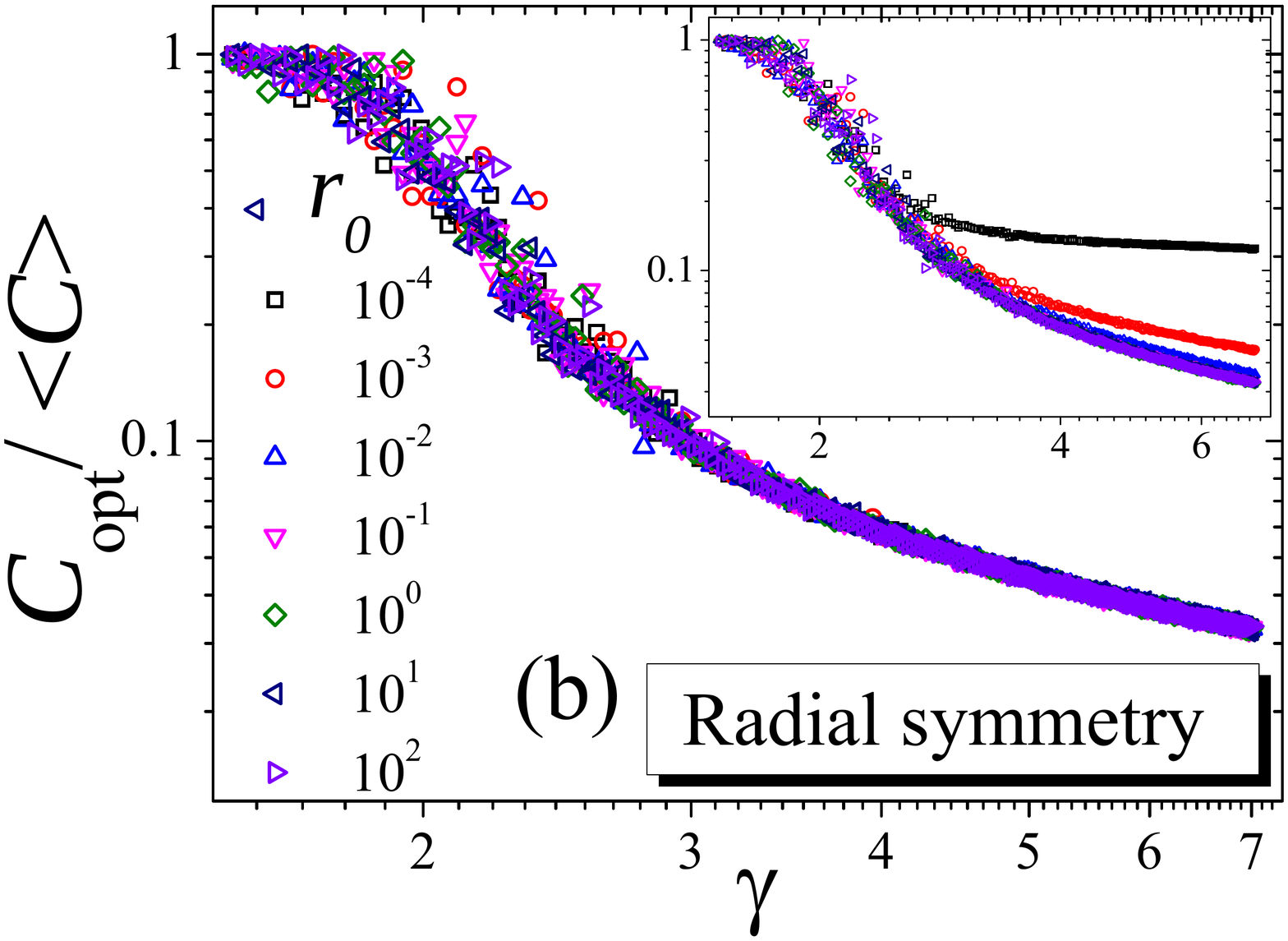}
\end{center}
\caption{\textbf{(a)} $C_{opt}/\left\langle C\right\rangle $ as function of $%
\protect\gamma $ for different values of $\protect\varepsilon $ for
2-opt-SAGCS. We perform the scaling $C_{s}=C/\protect\sqrt{(x_{\max
}-x_{\min })(y_{\max }-y_{\min })}$. The inset plot shows the same curves
without scaling. \textbf{(b)} The same plot by using power law coordinates
with radial symmetry (Eq. \protect\ref{Eq:Power_law_II}). }
\label{Fig:long_range_twist}
\end{figure}

Performing similar simulations to the plot of the Fig. \ref{Fig:long_range}
(a) but for the case 2opt-SAGCS, which can be observed in Fig. \ref%
{Fig:long_range_twist} (a) and the same conclusions can be drawn for $\gamma
<2$ -- the SA has no efficiency. In this case, SS-SAGCS and 2-opt-SAGCS are
obviously equivalent since the scenario is really complicate. In Fig. \ref%
{Fig:long_range_twist} (b) we show the case where we used power-law
coordinates with radial symmetry (Eq. \ref{Eq:Power_law_II}). Surprisingly,
the results do not change, showing that, independently from power law
coordinates, the phenomena of the performance of the SA is universal as
function of $\gamma $.

\section{Summary and Conclusions}

\label{sec:conclusions}

In this paper, we study the effects of the statistics on the coordinates of
the points when we apply an standard simulated annealing algorithm to the
travelling salesman problem. Our results are concerned with the long tail
effects on the coordinates but also the correlation effects between these
coordinates.

Our study shows that the performance of the simulated annealing increases as
the correlation increases in both versions of the SA. The main reason is the
dimensionality reduction, which transforms the simulated annealing at limit
in an approximated sorting algorithm. The correlation effects show that the
shape of distribution attributed to coordinates is more important than the
variance when we compare Gaussian and uniform distributions.

Our results also suggest that the higher the exponent of the power law, the
lower the simulated annealing performance, and for $\gamma <2$, i.e.,
distributions without the first moment, we have no performance of the SA,
i.e., $C_{opt}/\left\langle C\right\rangle \approx 1$ even when one
considers the benchmark for the SA, i.e., when the new configurations are
obtained with 2-opt.

We obtained an universal behavior of $C_{opt}/\left\langle C\right\rangle $ $%
\times \rho $ and $C_{opt}/\left\langle C\right\rangle $ $\times \gamma $
never explored in the optimization problems with SA. In addition, we also
show that 2-opt-SAGCS presents a better performance as larger is the system,
differently from its inefficient version, the SS-SAGCS, where the reverse
situation occurs.

We believe that both (correlation and long range effects) studies can bring
an interesting knowledge for more technical applications in artificial
intelligence, machine learning, and other areas. In special, in the search
for global minimum in neural networks algorithms, maybe reviving the
interests in simulating annealing as a viable alternative to stochastic
gradient descent at the optimization step.

\section*{Acknowledgments}

R. da Silva thanks CNPq for financial support under grant numbers
311236/2018-9, and 424052/2018-0. A. Alves thanks Conselho Nacional de
Desenvolvimento Cient\'{\i}fico (CNPq) for its financial support, grant
307265/2017-0. This research was partially carried out using the
computational resources from the Cluster-Slurm, IF-UFRGS. We would also like
to thank the anonymous referee for the excellent suggestions and
observations.

\section*{Appendix: Average distance between two points uniformly
distributed in a square}

\label{Section:Appendix}

Our original problem considers points uniformly distributed in the square
such that $x\in \lbrack -1,1]$, $y\in \lbrack -1,1]$, i.e., an square of
side 2. This is similar to consider points in the square defined by $x\in
\lbrack 0,2]$, $y\in \lbrack 0,2]$, and thus the average distance between
the points can be calculated by:

\begin{equation*}
\begin{array}{lll}
\left\langle d\right\rangle & = & \frac{1}{2^{4}}\int_{0}^{2}\int_{0}^{2}
\int_{0}^{2}\int_{0}^{2}\sqrt{(x_{i}-x_{j})^{2}+(y_{i}-y_{j})^{2}}
dx_{i}dx_{j}dy_{i}dy_{j} \\ 
&  &  \\ 
& = & \frac{1}{2^{3}}\int_{0}^{2}\int_{0}^{2}\int_{0}^{2}\int_{0}^{2}\sqrt{( 
\frac{x_{i}-x_{j}}{2})^{2}+(\frac{y_{i}-y_{j}}{2})^{2}}
dx_{i}dx_{j}dy_{i}dy_{j}%
\end{array}%
\end{equation*}

Performing the change of variables $z_{1}=\frac{x_{i}}{2}$, $z_{2}=\frac{
x_{j}}{2}$, $z_{3}=\frac{y_{i}}{2}$, and $z_{4}=\frac{y_{i}}{2}$ ,and thus

\begin{equation*}
\left\langle d\right\rangle
=2\int_{0}^{1}\int_{0}^{1}\int_{0}^{1}\int_{0}^{1}\sqrt{
(z_{1}-z_{2})^{2}+(z_{3}-z_{4})^{2}}dz_{1}dz_{2}dz_{3}dz_{4}
\end{equation*}

If $z_{1}$ and $z_{2}$ are uniformly distributed, so the random variable $%
u=\left\vert z_{1}-z_{2}\right\vert $ has a triangular distribution and the
probability density function $2(1-u)$, and thus making the change of
variables $u=\left\vert z_{1}-z_{2}\right\vert $, $v=\left\vert
z_{3}-z_{4}\right\vert $, $w=z_{2}$, and $t=z_{3}$, one obtains:

\begin{equation*}
\left\langle d\right\rangle =8\int_{0}^{1}\int_{0}^{1}(1-u)(1-v)\sqrt{
u^{2}+v^{2}}dudv
\end{equation*}

Now is almost done! Making $u=r\cos \theta $ and $v=r\sin \theta $. However
a trick is to perform the integration for the lower triangle where $0\leq
\theta \leq \pi /4$ 
\begin{equation*}
\begin{array}{lll}
\left\langle d\right\rangle & = & 16\int_{0}^{\pi /4}\int_{0}^{1/\sin \theta
}(1-r\cos \theta )(1-r\sin \theta )r^{2}drd\theta \\ 
&  &  \\ 
& = & 16\int_{0}^{\pi /4}\int_{0}^{1/\cos \theta }\left( r^{2}-r^{3}\left(
\sin \theta +\cos \theta \right) +r^{4}\cos \theta \sin \theta \right)
drd\theta \\ 
&  &  \\ 
& = & 16\left[ \frac{1}{3}\int_{0}^{\pi /4}\frac{1}{\cos ^{3}\theta }d\theta
-\frac{1}{4}\int_{0}^{\pi /4}\frac{\left( \sin \theta +\cos \theta \right) }{
\cos ^{4}\theta }d\theta +\frac{1}{5}\int_{0}^{\pi /4}\frac{\cos \theta \sin
\theta }{\cos ^{5}\theta }d\theta \right] \\ 
&  &  \\ 
& = & 16\left[ \frac{1}{12}\int_{0}^{\pi /4}\frac{1}{\cos ^{3}\theta }
d\theta -\frac{1}{20}\int_{0}^{\pi /4}\frac{\sin \theta }{\cos ^{4}\theta }
d\theta \right]%
\end{array}%
\end{equation*}

And finally by performing these integrals one obtains

\begin{equation*}
\left\langle d\right\rangle =\frac{1}{15}(4+2\sqrt{2}+10\ln (\sqrt{2}+1))
\end{equation*}

\bigskip

\end{document}